\documentclass{article}

\usepackage{PRIMEarxiv}

\usepackage[utf8]{inputenc}
\usepackage[T1]{fontenc}
\usepackage{makeidx}
\usepackage{url}
\usepackage{floatrow}
\usepackage{graphicx}
\usepackage{caption}
\usepackage{subcaption}
\usepackage[parfill]{parskip}

\usepackage{paralist}

\usepackage{amsmath}
\usepackage{amssymb}
\usepackage{amsthm}
\usepackage{mathtools}
\usepackage{bbm}
\usepackage{physics}
\usepackage{colonequals}
\usepackage[]{algorithm2e}
\usepackage{blkarray}
\usepackage[toc,page]{appendix}

\usepackage{breqn}

\usepackage{setspace}

\usepackage{tabularx}
\usepackage[table]{xcolor}
\newcolumntype{s}{>{\columncolor[HTML]{AAACED}} p{3cm}}

\usepackage{makecell}

\usepackage[maxbibnames=10, backend=biber]{biblatex}

\usepackage[bookmarks=true,colorlinks,pdfpagelabels,pdfstartview = FitH,bookmarksopen = true,bookmarksnumbered = true,linkcolor = black,plainpages = false,hypertexnames = false,citecolor = black,urlcolor=black]{hyperref}

\DeclareMathOperator*{\betadist}{Beta}
\DeclareMathOperator*{\dirichlet}{Dirichlet}

\DeclareMathOperator*{\categorical}{Cat}

\newcommand{\yes}{\texttt{yes}}
\newcommand{\no}{\texttt{no}}
\newcommand{\cs}{\texttt{cs}}

\newcommand{\qcs}{q^{\cs}}

\newcommand{\englanfz}[1]{``#1''}

\newcommand\numberthis{\addtocounter{equation}{1}\tag{\theequation}}

\title{No Need to Sacrifice Data Quality for Quantity: Crowd-Informed Machine Annotation for Cost-Effective Understanding of Visual Data}

\begin{document}

\pagenumbering{arabic} 

\begin{flushleft}
    {\LARGE No Need to Sacrifice Data Quality for Quantity: Crowd-Informed Machine Annotation for Cost-Effective Understanding of Visual Data} \\
    \bigskip
    Christopher Klugmann\textsuperscript{1,*},
Rafid Mahmood\textsuperscript{2},
Guruprasad Hegde\textsuperscript{3},
Amit Kale\textsuperscript{3},
Daniel Kondermann\textsuperscript{1}
\bigskip
\\
1 Quality-Match GmbH, Heidelberg, Germany
\\
2 University of Ottawa, Ottawa, Ontario, Canada
\\
3 Bosch Research, Bangalore, India
\\
\bigskip
* Correspondence to: \texttt{ck@quality-match.com}
\bigskip
\end{flushleft}

\begin{abstract}
Labeling visual data is expensive and time-consuming. Crowdsourcing systems promise to enable highly parallelizable annotations through the participation of monetarily or otherwise motivated workers, but even this approach has its limits. The solution: replace manual work with machine work. But how reliable are machine annotators? Sacrificing data quality for high throughput cannot be acceptable, especially in safety-critical applications such as autonomous driving. In this paper, we present a framework that enables quality checking of visual data at large scales without sacrificing the reliability of the results.
We ask annotators simple questions with discrete answers, which can be highly automated using a convolutional neural network trained to predict crowd responses. Unlike the methods of previous work, which aim to directly predict soft labels to address human uncertainty, we use per-task posterior distributions over soft labels as our training objective, leveraging a Dirichlet prior for analytical accessibility.
We demonstrate our approach on two challenging real-world automotive datasets, showing that our model can fully automate a significant portion of tasks, saving costs in the high double-digit percentage range. Our model reliably predicts human uncertainty, allowing for more accurate inspection and filtering of difficult examples. Additionally, we show that the posterior distributions over soft labels predicted by our model can be used as priors in further inference processes, reducing the need for numerous human labelers to approximate true soft labels accurately. This results in further cost reductions and more efficient use of human resources in the annotation process.
\end{abstract}
\section{Introduction}

In recent years, supervised learning has become central to computer vision, yielding impressive results and driving significant advancements. Yet, its effectiveness heavily hinges on access to labeled data, particularly crucial in deep learning for training robust models. \textit{Crowdsourcing}, a concept named by Howe \cite{howe2006rise}, has become a widely adopted method for data annotation, allowing for extensive parallelization. Typically, objects are labeled multiple times, and by aggregating responses, such as through majority voting, more reliable labels are generated for machine learning tasks \cite{sheng2008get}. Visual labels, such as bounding boxes or keypoints, are often \textit{not} redundantly created due to cost constraints, leading to a higher susceptibility to errors. Consequently, commercial providers have emerged to validate annotated data, employing human annotators to perform careful quality checks. One way to perform such a quality check is to ask the human elementary questions about the object of interest to determine through a chain of reasoning whether the visual annotation is correct according to the specifications of a label guide.

\begin{figure}
    \centering
    \includegraphics[width=\textwidth]{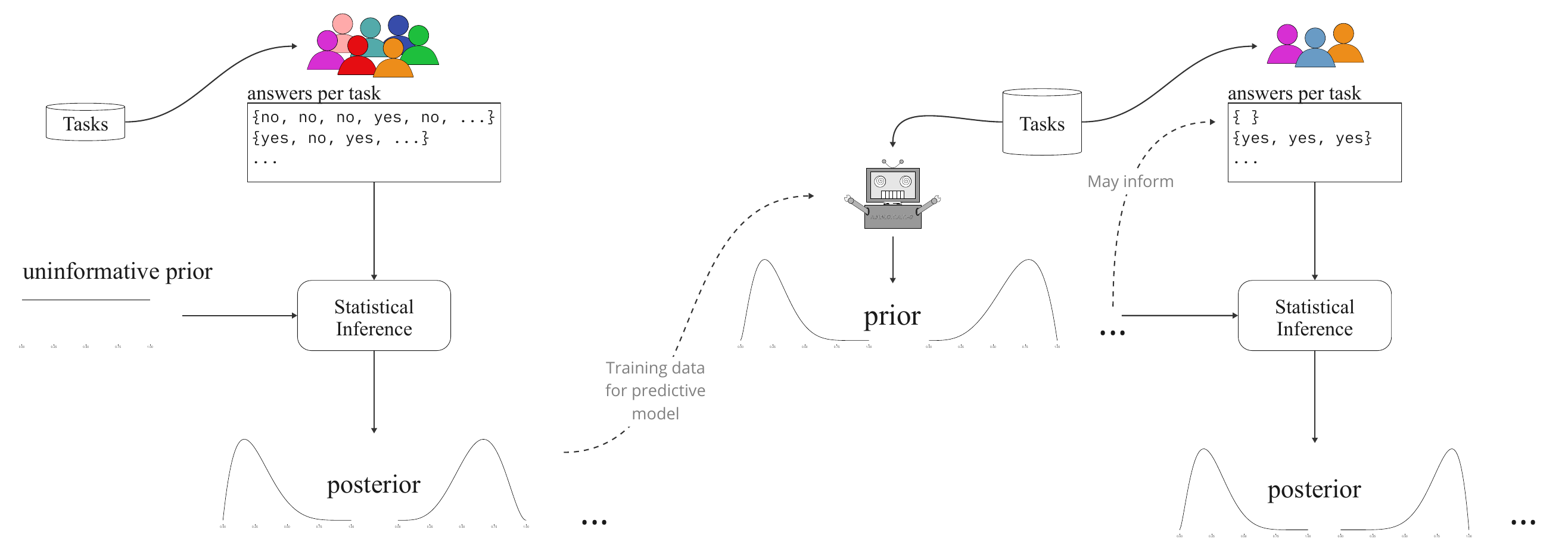}
    \caption{A crowd of labelers tackles categorical annotation tasks by providing sets of discrete answers per object. Bayesian inference is then applied to determine posterior distributions over task parameters. We train a machine to predict these \textit{actual posteriors}, which serve as priors for new tasks, reducing the need for extensive human input in reporting final task responses.}
    \label{fig:the-ominous-figure1}
\end{figure}

However, data quality checking, if done correctly, means having humans solve such simple mostly categorical annotation tasks in large numbers to maximize the chance of identifying actual label errors.
Even in a highly parallelizable and affordable framework such as crowdsourcing, such an approach faces limitations.
Consequently, both academia and industry have begun exploring \textit{auto-annotation} methods. Machine-guided annotation can entail aiding users in producing visual labels, such as proposing candidate boxes through object detection models, which users can then manually refine as necessary.
A good overview of different labeling strategies in computer vision and possibilities to automate labeling can be found in Sager et al.\ \cite{sager2021survey}. The aim of this work, however, is not to automate the generation of candidate labels, but to investigate how the subsequent quality inspection process can be supported in the context of a production ready system by using a model informed by human crowd responses. 

The goal of annotation is not to obtain mere unprocessed answers, but to make \textit{sense} of these answers. 
We propose training a model capable of autonomously solving categorical annotation tasks in large parts, while also informing the process of inferring latent ground truth from human worker responses for the remaining part of the data.
Figure \ref{fig:the-ominous-figure1} provides a high-level overview of this concept. We highlight the following contributions of our work:
\begin{itemize}
    \item Introducing a cost-effective process for inspecting visual data, featuring a model capable of predicting soft labels and distributions over them.
    \item Demonstrating the efficacy of our proposed model on two custom datasets (and various annotation questions) derived from publicly available sources, showcasing its ability to automate a significant portion of annotation tasks in the double-digit percentage range.
    \item Examining ambiguity in data labeling and demonstrating how it can be predicted to identify edge cases and difficult instances in visual datasets.
    \item Presenting a method for incorporating prior knowledge from the proposed model into the truth inference process in crowd sourcing, enriching human answers in the sense of Bayesian statistics.
\end{itemize}

\section{Related Work}

\paragraph{Truth Inference.} The central theme of our work is the process of data annotation and the inference of latent ground truth. A proven method to make sense of crowd workers' answers is to aggregate the answers appropriately. In the case of discrete responses, this can mean forming the majority response \cite{sheng2008get}. More elaborate methods exist that give the truth inference problem a probabilistic treatment, mostly inspired by the classic work of Dawid and Skene \cite{dawid1979maximum}, who used a variant of the Expectation Maximization Algorithm (EM) to infer latent variables such as error rates of labelers along with the ground truth. The original Dawid and Skene framework has long since been extended, for example to include the idea of estimating task difficulty as an additional latent parameter \cite{whitehill2009whose, ma2015faitcrowd}. Our work is agnostic to latent variables, such as those that can be used for modeling of tasks or workers. Instead, our objective is to prioritize analytically tractable data-generating processes, enabling us to leverage resulting distributions as learning targets for predictive models. Moreover, our focus is not solely on inferring \textit{hard labels} but rather on promoting the adoption of \textit{soft labels} — probability distributions over labels — as they offer a more nuanced and precise representation of the underlying characteristics of the objects or tasks in various scenarios.

\paragraph{Predictive Crowd Models.} Methods that not only consider the truth inference problem, but also consider subsequent learning from the inferred labels, are also mentioned in the literature. The work of Raykar et al.\ \cite{raykar2010learning} makes contributions in this regard by proposing a model that can make predictions for new unseen tasks. The estimation of ground truth taking into account latent parameters such as the specificity and sensitivity of the annotators in the context of a Bayesian approach is a by-product. A further development of this approach is the work of Rodrigues and Pereira \cite{rodrigues2018deep}, who train a neural network to learn the response behavior of a selected set of crowd workers in an end-to-end fashion. While this approach is similar to ours in terms of the predictive aspect, our focus is to learn on the transformed responses of the crowd without considering the identity of individual annotators (which is rarely possible for our setting). In this sense, our work is closer to the work of Wang et al.\ \cite{wang2017obtaining} who propose a predictive algorithm for task difficulty trained on empirical task difficulty gleaned from crowd worker responses. Predicting task difficulty enables more judicious allocation of annotation budgets, directing additional manual labeling efforts toward challenging tasks. This notion is further developed in \textsc{CrowdWT} \cite{tu2020crowdwt}, a fully Bayesian approach that simultaneously models task difficulty and annotator ability.

\paragraph{Soft Labels.} The seminal work by Peterson et al.\ \cite{peterson2019human} emphasizes direct learning from distributions resulting from solving visual tasks by multiple labelers, rather than relying solely on hard labels. 
Collins et al.\ \cite{collins2022eliciting} extend this concept, advocating for the use of \textit{per-annotator} probabilistic labels over soft labels derived from individual annotators' hard labels. Both studies illustrate that incorporating human uncertainty into training yields more robust models, as demonstrated on the \textsc{CIFAR-10} dataset \cite{krizhevsky2009learning}. While we also emphasize the importance of soft labels, our focus lies in predicting soft labels for automatic annotation within a production-ready system. Additionally, our approach distinguishes itself by considering distributions over soft labels as a natural outcome of a Bayesian methodology.
\section{A Simple Model of Annotation}

We aim to model the solving of \textit{discrete} annotation tasks under simple assumptions. Tasks are allocated to annotators $r$ from a pool of workers $[R] \equiv \{1, \dots, R\}$. Each task $t \in [T]$ involves answering a question about a single object, with the answer $a$ chosen from a set of discrete categories $\mathcal{C} = \mathcal{C}^{\prime} \cup \{ \cs \}$. We distinguish between \textit{proper} answer categories $\mathcal{C}^{\prime} \equiv [C]$ and an additional category, $\cs$ (\englanfz{can't solve}), indicating task \textit{unsolvability}.

The considered annotation tasks involve objects found in visual datasets. An annotator typically views an object representation (e.g., a section of an image highlighting the object of interest). Thus, a task comprises the object and the posed question. However, we condition the data-generating process on a fixed question, equating tasks and objects. We receive a sample $\mathcal{D} = \{( t_i, r_i, a_i)\}_{i=1}^N$ of $N$ answers during annotation. The notation implies no restrictions on the frequency with which an annotator tackles a task. However, the exact characteristics of individual annotators are not considered further in this paper. We treat all responses (conditioned on the respective task) as independent and identically distributed. We also assume that the annotation process is \textit{exhaustive}, i.e. every object $t \in [T]$ is solved by at least one annotator.

\subsection{Dirichlet-Multinomial Model}
\label{sec:dirichlet-multinomial-model}

By far the simplest model, and the one we choose in this paper, is the Dirichlet-multinomial model. We assign a probability vector $\vb{q}_t \in \Delta^C$ to each task $t \in [T]$, where $C = \abs{\mathcal{C}^{\prime}}$ denotes the number of proper categories and $\Delta^C$ the $C$-standard simplex (i.e. the set of discrete probability vectors over $C+1$ categories). The complete data-generating process is written as
\begin{align*}
    & \vb{q}_t \sim \dirichlet\left( \pmb{\alpha}_0 \right) \quad \forall \, t \in [T] && \text{(Prior probabilities)} \\
    & \pi_t = 1 - q^{\cs}_t \quad \forall \, t \in [T] && \text{(Solvability probability)} \\
    & p^k_t = q^k_t / (1 - q^{\cs}_t) \quad \forall \, t \in [T], \forall \, k \in \mathcal{C}^{\prime} && \text{(Conditional probability)} \\
    & a_i \sim \categorical\left( \pi_{t_i} p_{t_i}^1, \dots, \pi_{t_i} p_{t_i}^C, 1-\pi_{t_i} \right) \quad \forall \, i \in [N] && \text{(Likelihood)}
\end{align*}
Here we impose a Dirichlet prior on each $\vb{q}_t$. Typically, in the absence of better prior knowledge about each task, we choose $\pmb{\alpha}_0 = \pmb{1}_{C+1}$, i.e. a uniform prior. A remarkable and remarkably useful property of this simple population model is that the posterior distributions of $\vb{q}_t$ can be re-identified as Dirichlet distributions by exploiting the conjugacy of Dirichlet prior and multinomial likelihood. For each $t \in [T]$ we have
\begin{align}
    \vb{q}_t \, \vert \, \mathcal{D} \sim \dirichlet\left( \pmb{\alpha}_0 + \vb{n}_t  \right)\text{,}\label{eq:categorical_posterior}
\end{align}
where $\vb{n}_t = (n_t^1, \dots, n_t^C, n_t^{\cs})^{\intercal}$ denotes the frequencies with which the annotation of the task $t$ was decided for each of the categories.
We also note the existence of two transformed variables, the \textit{solvability probability} $\pi_t$ and the \textit{conditional probabilities} $\vb{p}_t$. These variables serve the purpose of adding interpretation to the components of each probability vector $\vb{q}_t$. They take into account the fact that for each task $t$ we are interested in two pieces of information, namely the general solvability of the task and the most probable outcomes, given that the task is found to be solvable. Using the change of variables formula for probability densities, the posterior marginal distributions can be written as
\begin{align}
    &\pi_t \, \vert \, \mathcal{D} \sim \betadist\left( \sum_{k=1}^C \alpha^k_t, \alpha^{\cs}_t\right)\text{,}\label{eq:transformed_dirichlet_pi}\\
    &\vb{p}_t \, \vert \, \mathcal{D} \sim \dirichlet\left( \alpha_t^1, \dots, \alpha_t^C\right)\text{,}
    \label{eq:transformed_dirichlet}
\end{align}
where $\pmb{\alpha}_t = \pmb{\alpha}_0 + \vb{n}_t$ is the parameter vector of the posterior Dirichlet distribution from \eqref{eq:categorical_posterior}. Note that for the special case $C=2$ (binary annotation task \englanfz{modulo} $\cs$) the conditional probability can be expressed by a single parameter, the (conditional) \textit{success probability}, which follows a beta distribution.

\paragraph{Point Estimation.}
The Bayesian framework allows direct probabilistic statements to be made about the probability vector $\vb{q}$ associated with a task.  However, it is also straightforward to derive a point estimator for $\vb{q}$ from the posterior distribution. A common choice is the \textit{posterior predictive probability}, denoted as $p(a \mid \mathcal{D})$, which represents the likelihood of observing a particular response $a$ given the already observed responses. If $\pmb{\alpha}$ is the parameter vector of the posterior distribution, then the probability vector associated with the posterior predictive probability distribution can be obtained as the expected value of $\vb{q}$ drawn from $\mathrm{Dirichlet}\left(\pmb{\alpha}\right)$.
The \textit{mode} of the posterior distribution serves as another point estimator, the one we consider. For Dirichlet distributions where $\alpha^k > 1$ holds for all $k$, the mode can be calculated according to
\begin{align}
    \hat{\vb{q}} \equiv \mathrm{mode}\left(\pmb{\alpha}\right) \equiv \left. \left(\pmb{\alpha} - \pmb{1} \right) \middle/ \sum_{k\in\mathcal{C}} \left( \alpha^k - 1\right) \right. \text{.}
    \label{eq:posterior_mode}
\end{align}
Note that the full information we obtain from Bayesian inference is represented by the entire posterior distribution, whereas an estimator like the one in equation \eqref{eq:posterior_mode} is merely a form of data reduction to study individual aspects of the posterior distribution. In the following, however, we will not only consider Dirichlet distributions that fall out of the true inference process, but also those that are predicted by a machine learning model. The estimator allows us to evaluate such a model in the light of an ordinary classification problem and to make comparisons, for example, with a classifier that has been trained to assign to each object the discrete label corresponding to the distinguished gold standard.

\paragraph{Why?}
Our interest is not so much in the simple population model, but in the applications that arise when we declare it to be the learning target of a predictive machine learning algorithm.
We imagine such a machine learning model as a black box that is able to predict the posterior distribution as it would emerge under the described inference process, given only the features of the task (i.e.\ the visual representation of the object). Such a predictive model can be seen as a further development of a mere classification model trained on hard labels. However, the proposed model is not just \textit{one} layer on top of the hard labels, which would mean a direct training on soft labels $\vb{q}$.
Instead, it is about predicting a \textit{distribution} over possible realizations of soft labels, which depends significantly on the number of observed responses and thus encodes uncertainty of any estimate regarding $\vb{q}$. We will shortly provide explanations of how all this comes together. First, however, we would like to briefly discuss how we measure ambiguity in response vectors and define what we mean by the distance between two probability vectors.

\subsection{Making Sense of Probability Vectors}

\paragraph{Ambiguity.}In the following, we are often confronted with the need to quantify features of discrete probability distributions (or \textit{soft labels}). The first measure we consider for this purpose is based on the idea of weighting two sources of label ambiguity and summarizing them in one statistic. For each outcome probability vector $\vb{q}$ we can calculate such a measure. To do this, we first recall the definition of solvability and conditional probabilities,
\begin{align}
    \pi = 1 - \qcs, \quad p^k = q^k / (1 - \qcs)\text{.}
\end{align}
Based on these quantities, we define the \textit{ambiguity} according to
\begin{align}
    \mathrm{ambiguity}\left(\vb{q}\right) = \begin{cases}
        1 - \frac{\eta}{2}\frac{C}{C-1} \sum_{k\in\mathcal{C}^{\prime}} \abs{p^k - \frac{1}{C}} &\text{if}\,\sum_{k\in\mathcal{C}^{\prime}} q^k > 0 \\
        1 &\text{else,}
    \end{cases}
    \label{eq:def-ambiguity}
\end{align}
where $\eta = \eta(\pi)$ is a scaling factor that depends only on the solvability of the task under consideration.
In this paper we make the choice of an exponential scaling factor $\eta(\pi) = \exp\left(\gamma [1-\pi] \right)$, where $\gamma$ is a hyperparameter.
We select $\gamma$ in such a way that the scaling factor assumes the value $\eta_0 = 0.4$ for a solvability of $\pi_0 = 0.8$. The exponential scaling reflects our conviction that even a small proportion of $\cs$ responses generally provides strong evidence for the ambiguity of the labeling task in question, more so than a linear scaling would express.
The intuition behind the ambiguity measure is that ambiguity or disambiguity is decided based on the distance of the conditional probability distribution $\vb{p}$ to a uniform distribution in terms of an $\ell_1$-like distance. The scaling factor modifies this base disambiguity by also taking into account the solvability or unsolvability of the task. In the case that the task is found to be completely unsolvable according to the observed probability distribution, the ambiguity is maximal, i.e.\ equal to 1. This measure, although simple, has practical relevance and can be used to filter crowd-annotated (or predicted) datasets and identify potentially difficult examples.

\paragraph{Confidence.}
Another property of discrete probability distributions that we want to quantify is the confidence of the predicted majority label. There are several ways to give such a confidence indication, but a simple and common one is based on looking at the score of the predicted label considered most likely under the distribution. We follow this approach, but min-max scale this score to obtain values that can take any value in the interval $[0, 1]$. For any discrete probability distribution over $C+1$ classes, the confidence (of the predicted majority label) is defined as
\begin{align}
    \mathrm{conf}(\vb{q}) = \frac{1}{C} \left\{ (C+1) \cdot \max_{k\in\mathcal{C}} q^k - 1\right\}\text{.}
    \label{eq:def_confidence}
\end{align}
If $\vb{q}$ is the prediction of a classification model trained on hard labels, there is a risk that the scores are poorly calibrated \cite{guo2017calibration, wei2022mitigating, hendrycks2021unsolved}. In our case, however, the distributions come from the prediction of a model trained on soft labels and thus approximate the true sentiment of a crowd of labelers.

\paragraph{Distance.}
Finally, we introduce a distance function that measures the distance of a predicted distribution $\hat{\vb{q}}$ to a reference distribution $\vb{q}$. We opt for the function
\begin{align}
    D(\hat{\vb{q}}, \vb{q}) = \max_{k\in\mathcal{C}} \frac{\abs{\hat{q}^k - q^k}}{\max\{q^k, 1 - q^k\}}
    \label{eq:def_softdist}
\end{align}
here, firstly because it provides interpretable values, and secondly because it has the desirable property that $D=1$ if $\hat{\vb{q}}$, $\vb{q}$ are one-hot for different categories.
Conversely, $D$ is equal to $0$ if and only if the predicted and actual soft label match exactly.
Another example could be that annotators solve a task unanimously, expressed by the soft label $\vb{q} = \begin{pmatrix} 0 & 1 & 0 \end{pmatrix}^{\intercal}$ over the categories $\texttt{no}$, $\texttt{yes}$ and $\texttt{cs}$. A machine learning model, on the other hand, could evaluate the corresponding object as slightly ambiguous and predict a soft label such as $\hat{\vb{q}} = \begin{pmatrix} 0.03 & 0.9 & 0.07 \end{pmatrix}^{\intercal}$. In this case, the distance $D$ is the largest of the three values $0.03/1$, $0.1/1$ and $0.07/1$, i.e.\ it is $D = 0.1$.

\section{Predicting Posterior Distributions}

\label{sec:predicting_posterior}

Whenever we use a Bayesian model as the basis for crowd inference, we must make a choice regarding the prior, i.e., answer the question of what prior knowledge is allowed to enter the process of statistical inference. So far, we have described the prior distributions as uninformative, i.e., not tending to one answer or the other, and this is, in light of what we can assume, a reasonable initial guess. However, in our case, there is unused data available for each task that can help to arrive at a better initial estimate. We can use the visual representations associated with the tasks under consideration to arrive at a better, a \textit{learned}, prior distribution for each task. To do this, we apply a process of splitting a dataset to train a predictive model on a small portion of the dataset, which will in turn generate improved prior distributions for the remaining portion of the data.

\begin{figure}
    \centering
    \includegraphics[width=\textwidth]{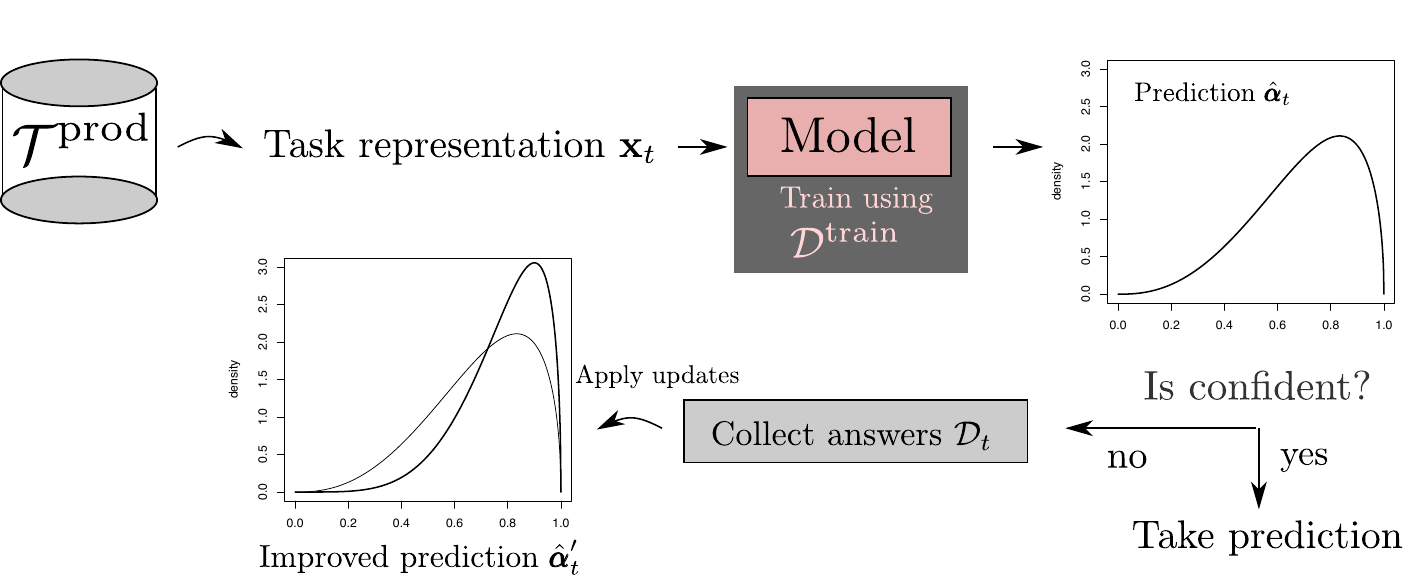}
    \caption{Overview of the process for generating new annotated data from a set of given unannotated image data. A model pre-trained on a small set of data sees unlabeled images for which it predicts the crowd response. If the confidence of the model prediction exceeds a certain threshold, the prediction can be used as ground truth and no further annotation is needed. Otherwise, the prediction is used as a prior distribution for further Bayesian updates.}
    \label{fig:dataprocess}
\end{figure}

\paragraph{The Data Process.} In Figure \ref{fig:dataprocess} we present the process we propose for semi-automated annotation of visual data.
The lower part of the figure contains the steps we described in the last sections, i.e., applying Bayesian updates to a prior distribution once we observe responses from labelers. The novelty in this process is that we no longer use an \textit{uninformative} prior distribution that ignores the image contents, but a \textit{learned} prior predicted by a blackbox model. Such a model is trained in advance on a portion of the total data to be annotated, evaluated, and later used as a predictive engine to generate informed prior distributions on the unseen and unannotated portion of the data. The process, from obtaining the raw data, i.e., image data to be annotated, to generating the labels on the entire dataset, is divided into the following steps.
\begin{enumerate}
    \item The unannotated dataset, consisting of visual data, is partitioned into a \textit{training} dataset, a \textit{development} dataset and a \textit{production} dataset. We call the corresponding index sets $\mathcal{T}^{\mathrm{train}}$, $\mathcal{T}^{\mathrm{dev}}$ and $\mathcal{T}^{\mathrm{prod}}$. The development dataset can be further decomposed into a \textit{validation} and \textit{test} dataset $\mathcal{T}^{\mathrm{val}}$ and $\mathcal{T}^{\mathrm{test}}$.
    \item We let all tasks from $\mathcal{T}^{\mathrm{train}}$ and $\mathcal{T}^{\mathrm{dev}}$ be annotated by a large number of human labelers to obtain annotated datasets $\mathcal{D}^{\mathrm{train}}$ and $\mathcal{D}^{\mathrm{dev}}$.
    \item We train a blackbox model on the training dataset $\mathcal{D}^{\mathrm{train}}$. We first infer for each task the corresponding posterior distribution in analytic form according to equation \eqref{eq:categorical_posterior}. The posterior distributions (or the parameters characterizing these distributions) are the learning target.
    \item Based on the development dataset, we choose a threshold that allows us to distinguish predictions of the model that are correct with high probability from those that are prone to error.
    \item Let the model make predictions on the visual representations corresponding to the tasks in $\mathcal{T}^{\mathrm{prod}}$, i.e., predict the corresponding Dirichlet distributions for each task. If a predicted distribution implies a higher certainty than the previously chosen threshold, we adopt the predicted response without requesting further annotations for this data. Let this set of tasks be $\mathcal{T}_{\mathrm{p}}^{\mathrm{prod}} \subseteq \mathcal{T}^{\mathrm{prod}}$.
    \item All remaining tasks $\mathcal{T}^{\mathrm{prod}} \setminus \mathcal{T}_{\mathrm{p}}^{\mathrm{prod}}$ we let annotate manually, applying Bayesian updates to the distribution predicted by the model.
\end{enumerate}
We perform the data split in such a way that only a small portion of a typically large dataset is used to train and validate the actual model.
The development dataset referenced here is understood as the union of the validation and test datasets. On the validation dataset, we automatically determine a confidence threshold depending on a declared target quality of the model predictions. If available, the further generalization ability of the model and the choice of threshold can be assessed on the test dataset. Since the training and development dataset together are assumed to be small (relative to the production dataset), this part of the data can be annotated by humans in a large number of repetitions per task to get a reliable sentiment picture of the crowd.
From these responses we derive posterior distributions, as we would otherwise report in the classical process without model guidance. The information inferred on the training dataset helps us to predict better distributions outside the seen dataset, i.e. for the large fraction of unannotated data, which technically correspond to the expected posterior distribution, but are used here as prior for further Bayesian updates. A significant advantage of this method is that we do not need to further consider simple tasks in the production set after a few responses have been received. Some of these tasks, in general even a very large part, are predicted so reliably by the blackbox model that in these cases we always declare the model response to be the \englanfz{ground truth}. This saves resources, i.e. human manpower, which we can use more sensibly for more difficult tasks. Details on how we identify particularly simple tasks that we no longer need to manually annotate follow later.

\paragraph{Feature Representation.}
We now turn to the question of what a model looks like that predicts the crowd-induced per-task probability distribution. Since we work with image data, it is an obvious requirement to use such models that can learn strong visual representations. To this end, we choose as a backbone for our model a convolutional neural network (CNN) that adapts one of the well known architectures from the literature. We choose to use a \textsc{ResNet-50} \cite{he2016deep} feature extractor in our experiments, however any other model that learns a strong visual representation of the underlying data will work. Since in crowd learning we typically want to use as little data as possible to predict crowd responses at minimal cost, we always base parameter inference on the weights of a model pre-trained on a significantly larger dataset. We follow the usual approach of using a model pre-trained on the \textsc{ILSVRC 2012 ImageNet} dataset \cite{russakovsky2015imagenet}, which we finetune end to end as part of our crowd learning.

Unlike for ordinary image classification, when annotating data we are mostly interested in identifying specific objects in specific image regions. What sounds like object detection differs from it in that we assume the localization of relevant objects is known. Our goal is to use the wisdom of the crowd to test a set of potentially interesting objects for certain semantic properties using simple, categorical questions. Thus, we remain in the framework of discrete classification, enriched by information of the object under consideration. We therefore allow the model to consider binary masks as input, in addition to crops of images. The mask is processed by a single convolutional layer and the activations are additively merged at an early level of the network with the activation of the image branch. We follow the literature and initialize the weights of the masking branch with zeros to avoid biasing the original feature extraction of the pre-trained network \cite{eppel2018classifying}.

\paragraph{Model Output.}
For each task object, the crowd learning model can make a prediction about the Dirichlet distribution associated with the task and the considered crowd of labelers. The feature representation of the task, computed as described in the previous paragraph, is transformed into an output $\vb{z} \in \mathbb{R}^{C+1}$ through a single fully connected layer. These \englanfz{raw} outputs must be converted into parameters of a Dirichlet distribution. We want to allow for the possibility that the Dirichlet distributions on which the model is trained are inferred for different numbers of observed responses. For this reason, we consider the number of observed responses per task as additional input to the machine learning model. For a true posterior Dirichlet distribution, we know that the sum of the parameters must be equal to the sum of the prior parameters added to the number of observed responses. Let the number of observed responses be denoted by $n$ and the prior parameter sum by $\alpha_0^0$. The final parameters predicted by the model then result as
\begin{align*}
    &\hat{\pmb{\alpha}} = \alpha_0^0 \vb{s} + n\pmb{\sigma}, \text{where} \numberthis \label{eq:dirichlet_scaling}\\
    & \vb{s} = \mathrm{Softmax}\left(\vb{z} \right) \text{ and }\\
    & \pmb{\sigma} = \mathrm{Softmax}\left(\vb{W}\vb{z} \right) \text{.}
\end{align*}
Here, $\vb{W} \in \mathbb{R}^{C+1 \times C+1}$ is a matrix of parameters that are learned together with the other parameters of the network. The resulting vector $\hat{\pmb{\alpha}}$ has the desirable property that $\sum_{k}\hat{\alpha}^k = \alpha_0^0 + n$, just like the \englanfz{actual} posterior distribution. During inference time, the number of responses $n$ on which the model prediction is to be based can be fixed, e.g.\ as a value that is representative of the setting under consideration.

\paragraph{Training Objective.}

The goal of crowd learning is to predict probability distributions, one per task. Measures of the difference between two probability distributions, have been studied in the context of information geometry. A common choice is the Kullback-Leibler divergence, but it is known to be inappropriate for Dirichlet distributions \cite{rauber2008probabilistic}, which is what we are interested in. We use instead the Chernoff distance \cite{chernoff1952measure}, more precisely the Bhattacharyya distance, which we minimize between the predicted and actual distributions over the observed data.
For two $d$-dimensional probability distributions coming from a family of distributions $p(\,\cdot\,; \pmb{\alpha})$, the Chernoff distance is defined by
\begin{align}
    J(\pmb{\alpha}_a, \pmb{\alpha}_b) = -\log \int p^{\tau}\left( \vb{x}; \pmb{\alpha}_a\right) p^{1-\tau}\left( \vb{x}; \pmb{\alpha}_b\right) \dd^d x\text{,}
    \label{eq:chernoff}
\end{align}
where $\tau \in (0, 1)$. For $\tau=1/2$ the Bhattacharyya distance is obtained.
If $p(\, \cdot \, ; \pmb{\alpha})$ is the class of Dirichlet distributions of order $C$ and thus $\pmb{\alpha}$ elements of the $C$-dimensional standard simplex, then the Chernoff distance \eqref{eq:chernoff} can be calculated analytically according to
\begin{dmath}
    J(\pmb{\alpha}_a, \pmb{\alpha}_b) = \ln \Gamma \left( \sum_{k=1}^C \left( \tau \alpha^k_a + (1- \tau) \alpha_b^k\right)\right)
    + \tau \sum_{k=1}^C \ln \Gamma\left( \alpha_a^k\right)
    + (1-\tau) \sum_{k=1}^C \ln \Gamma\left( \alpha_b^k\right)
    - \sum_{k=1}^C \ln \Gamma \left( \tau \alpha^k_a + (1- \tau) \alpha_b^k\right)
    - \tau \ln \Gamma \left( \norm{\pmb{\alpha}_a}_1 \right) - (1 - \tau) \ln \Gamma\left( \norm{\pmb{\alpha}_b}_1 \right)\text{,}
    \label{eq:chernoff_analytical}
\end{dmath}
where $\norm{\vb{v}}_1$ denotes the 1-norm of a vector $\vb{v}$. Using this analytical expression for the Chernoff distance, we can define a loss function for our crowd learning model and optimize it within the backpropagation framework using mini-batch based stochastic gradient descent updates.
\section{Datasets}

Our work aims to provide actionable insights and present researchers and industry practitioners with tools for exploring visual data at large scales. Our approach is based on accelerating data inspection by a machine that learns from distributions of human workers' responses.
While benchmark datasets like \textsc{CIFAR10H} \cite{peterson2019human, battleday2020capturing} are prevalent in distribution learning, our focus in this paper shifts towards alternative datasets for method evaluation. In pursuit of assessing our approach against datasets that more authentically capture the challenges of real-world annotation processes, we introduce two custom datasets situated within the automotive domain. These datasets, derived from publicly available sources, address two significant use cases: the detection of vulnerable road users (VRUs) in traffic scenes and the identification of traffic signs. Both classes encompass diverse objects, ranging from pedestrians to motorcyclists or cyclists within the VRU category.
We reiterate the fact that the goal of our proposed (auto-)annotation process is to semantically enrich datasets. What is learned about the objects of interest through human or machine annotation can be used to filter the dataset in question, e.g. with the aim of excluding less relevant, uninformative or even \textit{false-positive} objects in downstream applications such as the training of a detection model.

\subsection{Do you see a \textit{human being}?}

The first dataset we consider as part of our experiments is derived from the widely cited \textsc{EuroCity Persons} (ECP) dataset \cite{braun2019eurocity}.
The original ECP dataset contains road traffic scenes recorded in different cities within Europe with over 238k annotated VRUs such as pedestrians or cyclists. In the course of our work, we use a subset of the daylight imagery of the dataset, which is \textit{re-annotated} by a well-known label service with respect to the VRUs within these images. Although details about the annotation process of the ECP dataset are not exactly known to us, we think that our annotation differs significantly from that of the original authors in terms of both approach and taxonomy. We do so motivated by the thought that such a process, as commercially pushed and offered by labeling platforms, is closer to the reality lived by machine-learning practitioners. Our claim is to operate on off-the-shelf data that can be improved by our proposed lightweight and semi-automated curation process to develop trustworthy intelligent systems faster and cheaper.

In total, we let the label provider examine a subset of 5077 randomly chosen frames from the set of ECP street scenes. As a result, we obtain a dataset with 2d bounding box annotations of 32711 objects. These objects are which we want to check for the correctness of the \englanfz{human being} label through our own annotation pipeline. For each of the 32711 tasks, we let workers answer the question whether they see a human being in a 2d bounding box rendered on a section of a frame. We run this process exactly 20 times for each task, obtaining a total of 20 labels per task, which are chosen from the set of allowed answers $\{\no, \yes, \cs \}$. It is worth mentioning that the number of active raters is large and for every two tasks the intersection of raters working on the respective task is generally small.
Furthermore, these tasks are addressed by a crowd of gamers who engage in our annotation process in exchange for in-game currency through an online gaming platform. Ewecker et al.\ \cite{ewecker2024} recently evaluated and contrasted the efficacy of this annotation model with traditionally monetarily incentivized crowds comprising \englanfz{professional annotators}.

\subsection{Traffic Sign Attributes}

The other dataset we use for our experiments is a subset of the \textsc{Mapillary Traffic Sign Dataset} (MTSD) \cite{ertler2020mapillary}, which contains diverse imagery from road traffic scenes around the world.
The original dataset contains 100k high-resolution images with traffic signs fully or partially annotated by bounding boxes. The dataset uses a comprehensive taxonomy of 300 traffic sign classes. For our application, we use the original bounding box annotations of the dataset as-is, but neglect the taxonomy, i.e., treat objects class-agnostically and attempt to infer semantic attributes through crowd sourcing.
This is part of an originally broader data pipeline, within which an attempt is made to extract the relevant (i.e., true-positive) annotations from a large number of candidate annotations.
We also use only a subset of 3957 of the total available imagery, which we select randomly. We use the imagery and its annotations in an essentially unmodified way, but perform an additional false negative check, i.e., we augment the dataset with such bounding boxes that were missed by the authors of the dataset. This results in a dataset containing a total of 21678 objects.


\begin{table}
\centering
\footnotesize
\begin{tabularx}{\textwidth}{m{2.5cm}m{3cm}X}
    \Xhline{0.5\arrayrulewidth}
    & question & answer choices \\
    \Xhline{0.5\arrayrulewidth}
    occluded & Is the sign occluded? & no, yes \\
    \Xhline{0.25\arrayrulewidth}
    heavily-occluded & Is the sign heavily occluded? & no, yes \\
    \Xhline{0.25\arrayrulewidth}
    border & Does the sign have a border? & no, yes \\
    \Xhline{0.25\arrayrulewidth}
    shape & What is the shape of the sign? & rectangle, diamond, triangle, octagon, circle, other \\
    \Xhline{0.25\arrayrulewidth}
    side & Which side of the sign can be seen? & back of traffic sign, back of regular sign, front of traffic sign, front of regular sign \\
    \Xhline{0.25\arrayrulewidth}
    condition & What is the condition of the sign? & rotated, blurry, dirty, damaged, good conditions
\end{tabularx}
\caption{The attributes considered for the Mapillary dataset together with their possible values, i.e. permissible answer categories (except for \englanfz{can't solve}). In addition, the wording of the question shown to the annotators is given per attribute.}
\label{table:butternut_questions}
\end{table}

Unlike in the case of the ECP dataset, for the objects of the Mapillary dataset we are interested in answering six individual attribute questions, which are listed in table \ref{table:butternut_questions} together with the allowed answers (except for \englanfz{can't solve}) from which the annotators choose. Another difference from the previous dataset is that we no longer ask gaming crowds but professional annotators to solve the annotation tasks for us. These crowds are characterized by the fact that individuals typically have experience in solving annotation tasks and do so at an increased cost compared to the gaming crowd. We label each task with five repetitions, i.e.\ for each pair of object and question we receive exactly five labels from the paid crowd workers.
\section{Experiments}

\subsection{Setup}

\begin{figure}
    \centering
    \includegraphics[width=\textwidth]{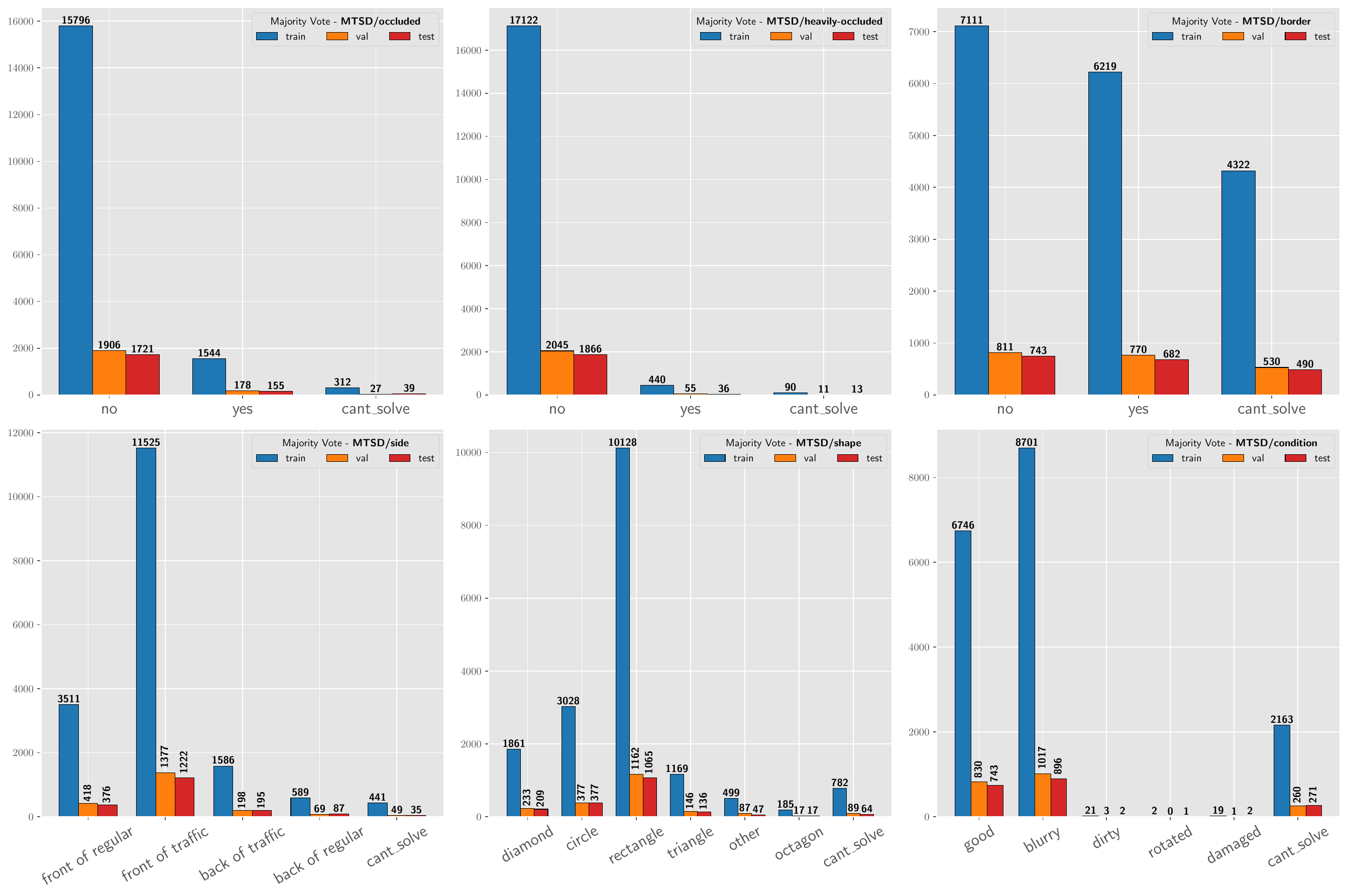}
    \caption{Distribution of majority vote within each attribute of the Mapillary dataset, stratified by data split. The numbers above the bars indicate the absolute frequencies with which the majority answers were selected for the respective annotation questions. For most of the datasets considered, the distribution of responses is highly non-uniform, i.e.\ majority responses typically fall into a few categories, while others are strongly underrepresented.}
    \label{fig:butternut_mv_dist}
\end{figure}

\paragraph{Datasplit.}
For training and evaluation of the machine learning models, we split the datasets into training, test and validation datasets based on the respective image frames according to an 80-10-10 split. Due to the coarse sampling, we expect the data used for training to be different enough from the data on which we evaluate the models. Exact details of the data split for the ECP and Mapillary data can be found in tables \ref{table:ecp_split} and \ref{table:butternut_split} respectively. In the case of the ECP dataset, we have also provided an indication of the distribution of majority responses per subset. The same, but in graphical form, is shown in Figure \ref{fig:butternut_mv_dist} for the individual attributes of the Mapillary data. We would like to emphasize that the distributions of the majority responses can only give a rough indication of the content of a crowd-annotated dataset and leave many details such as entropy of the soft labels inaccessible.

\begin{table}
\centering
\footnotesize
\begin{tabularx}{\textwidth}{m{0.5cm}m{0.5cm}m{0.5cm}m{2cm}!{\vrule width 0.25\arrayrulewidth}m{0.5cm}m{0.5cm}m{0.5cm}}
\Xhline{0.25\arrayrulewidth}
 & \#frames & \#boxes & \#boxes/frame & \multicolumn{3}{c}{\textsc{MV} distribution} \\
 &  &  &  & \texttt{no} & \texttt{yes} &  \cs \\
\Xhline{0.25\arrayrulewidth}
 train & 4029 & 26080 & 6.47 & 2096 & 21793 & 2191 \\
 val & 500 & 3158 & 6.32 & 256 & 2610 & 292 \\
 test & 548 & 3473 & 6.34 & 294 & 2889 & 290 \\
\end{tabularx}
\caption{Dataset split statistics for the ECP-derived dataset. For each subset, the table displays the number of images, bounding boxes, and the average number of boxes per image. Additionally, the distribution of majority responses within each subset is provided.}
\label{table:ecp_split}
\end{table}

\begin{table}
\centering
\begin{tabularx}{\textwidth}{cccc}
    \Xhline{0.25\arrayrulewidth}
    & \#frames & \#boxes & \#boxes/frame \\
    \Xhline{0.25\arrayrulewidth}
    train & 3165 & 17652 & 5.58 \\
    val & 395 & 2111 & 5.34 \\
    test & 397 & 1915 & 4.82
\end{tabularx}
\caption{Dataset split statistics for the Mapillary-derived dataset.}
\label{table:butternut_split}
\end{table}

\paragraph{Models.}

We train three types of models for our experiments: our proposed Dirichlet model, along with two baseline models labeled \texttt{dirichlet}, \texttt{baseline\_hard}, and \texttt{baseline\_soft} respectively. All models share a common architecture as described in Section \ref{sec:predicting_posterior}, comprising a \textsc{ResNet-50} backbone \cite{he2016deep} pretrained on \textsc{ImageNet} \cite{russakovsky2015imagenet}, augmented by a masking branch to indicate the object's position within the image crop.
Additionally, we experimented with an \textsc{EfficientNet} \cite{tan2019efficientnet} but observed negligible differences, hence we omit its reporting. The CNN-generated feature representation is passed through a single fully-connected layer with output units corresponding to the classes, including the \englanfz{can't solve} category. All models share the same architecture, although in the case of the Dirichlet model we consider an additional weight matrix $\vb{W}$, as shown in equation \eqref{eq:dirichlet_scaling}.

For the baseline models, we employ an ordinary cross-entropy loss function. In \texttt{baseline\_hard}, the loss is computed between the softmax-activated model predictions and one-hot distributions derived from majority voting of crowd responses. In contrast, \texttt{baseline\_soft} is trained directly on label distributions, akin to Peterson et al.\ \cite{peterson2019human}. For the Dirichlet model, we use the Bhattacharyya distance between predicted and actual parameters of the posterior Dirichlet distribution (see equation \eqref{eq:chernoff_analytical}). Given the unbalanced nature of the datasets we study, we introduce class weights for all loss calculations. When training on hard labels, the weight assigned to an object receiving the majority vote for class $y$ is determined as:
\begin{align}
w(y) = \left. (T + \abs{\mathcal{C}}) \middle/ (\abs{\mathcal{C}} \cdot (T_y + 1))\right.\text{,}
\end{align}
where $T$ denotes the total number of training examples and $T_y$ represents the number of examples for class $y$.
For \texttt{baseline\_soft} and \texttt{dirichlet}, we employ heuristically motivated loss weights. Specifically, a training example associated with the \englanfz{real} soft label $\vb{q}$ is incorporated into training using a convex combination of corresponding hard label weights,
\begin{align}
\widetilde{w}(\vb{q}) = \sum\nolimits_y q^y w(y)\text{.}
\label{eq:soft_label_weights}
\end{align}
All models are implemented in \textsc{Pytorch} \cite{paszke2019pytorch}.
The hyperparameters are chosen identically for all models in a way that the values are close to default configurations, with only small adjustments.
We utilize the \textsc{Adam} Optimizer \cite{kingma2014adam} with a learning rate of $2\mathrm{e}{-4}$, $\beta_2 = 0.995$ and default values for other parameters. To facilitate end-to-end training, we adopt a linear learning rate warm-up strategy over approximately $2/(1-\beta_2)$ iterations, following Ma and Yarats' recommendation \cite{ma2021adequacy}.

The models expect cropped images (and masks) as input, typically square and resized to 
$224\times 224$ pixels. Training is performed on a single \textsc{NVIDIA GeForce RTX 4090}, with a batch size of $256$. We refrain from using data augmentations and extensive hyperparameter sweeps, as the primary aim of this study is to showcase the efficacy of the proposed framework rather than achieving peak accuracy.

\subsection{Hard and Soft Model Comparison}

\begin{table}[htbp]
\centering
\footnotesize
\begin{tabularx}{\textwidth}{m{1.2cm}XXXXXXX}
\Xhline{0.5\arrayrulewidth}
\multicolumn{8}{l}{\textbf{ECP/human-being}} \\
\textbf{} & $\mathrm{acc}$ & $\mathrm{prec}_{\cs}$ & $\mathrm{rec}_{\cs}$ & $\langle D \rangle$ & $\langle D\rangle_{\widetilde{w}}$ & $\langle H \rangle$ & $\langle H\rangle_{\widetilde{w}}$ \\
\Xhline{0.25\arrayrulewidth}
\textbf{\texttt{hard}} & 89.3\% & 48.1\% & \textbf{67.2\%} & 0.253 & 0.522 & 0.941 & 1.872  \\
\textbf{\texttt{soft}} & 91.7\% & 60.6\% & 58.3\% & 0.138 & 0.254 & 0.519 & 0.973  \\
\textbf{\texttt{dirichlet}} & \textbf{91.8\%} & \textbf{61.3\%} & 55.2\% & \textbf{0.132} & \textbf{0.245} & \textbf{0.515} & \textbf{0.968}  \\
\Xhline{0.5\arrayrulewidth}
\multicolumn{8}{l}{\textbf{MTSD/occluded}} \\
\textbf{} & $\mathrm{acc}$ & $\mathrm{prec}_{\cs}$ & $\mathrm{rec}_{\cs}$ & $\langle D \rangle$ & $\langle D\rangle_{\widetilde{w}}$ & $\langle H \rangle$ & $\langle H\rangle_{\widetilde{w}}$ \\
\Xhline{0.25\arrayrulewidth}
\textbf{\texttt{hard}} & 90.6\% & 43.3\% & \textbf{33.3\%} & 0.150 & 0.552 & 0.471 & 1.795  \\
\textbf{\texttt{soft}} & 90.8\% & 25.0\% & 2.6\% & 0.162 & 0.449 & 0.341 & 1.102  \\
\textbf{\texttt{dirichlet}} & \textbf{92.7\%} & \textbf{66.7\%} & 30.8\% & \textbf{0.147} & \textbf{0.439} & \textbf{0.327} & \textbf{1.079}  \\
\Xhline{0.5\arrayrulewidth}
\multicolumn{8}{l}{\textbf{MTSD/heavily-occluded}} \\
\textbf{} & $\mathrm{acc}$ & $\mathrm{prec}_{\cs}$ & $\mathrm{rec}_{\cs}$ & $\langle D \rangle$ & $\langle D\rangle_{\widetilde{w}}$ & $\langle H \rangle$ & $\langle H\rangle_{\widetilde{w}}$ \\
\Xhline{0.25\arrayrulewidth}
\textbf{\texttt{hard}} & 97.5\% & \textbf{100\%} & 7.7\% & \textbf{0.066} & 0.725 & 0.229 & 2.690  \\
\textbf{\texttt{soft}} & 97.7\% & - & 0.0\% & 0.077 & 0.589 & 0.170 & 1.586  \\
\textbf{\texttt{dirichlet}} & \textbf{97.9\%} & \textbf{100\%} & \textbf{15.4\%} & 0.069 & \textbf{0.564} & \textbf{0.159} & \textbf{1.483}  \\
\Xhline{0.5\arrayrulewidth}
\multicolumn{8}{l}{\textbf{MTSD/condition}} \\
\textbf{} & $\mathrm{acc}$ & $\mathrm{prec}_{\cs}$ & $\mathrm{rec}_{\cs}$ & $\langle D \rangle$ & $\langle D\rangle_{\widetilde{w}}$ & $\langle H \rangle$ & $\langle H\rangle_{\widetilde{w}}$ \\
\Xhline{0.25\arrayrulewidth}
\textbf{\texttt{hard}} & 86.9\% & 84.4\% & \textbf{86.0\%} & 0.232 & 0.692 & 0.648 & 2.497  \\
\textbf{\texttt{soft}} & 86.8\% & 84.6\% & 85.2\% & 0.233 & \textbf{0.634} & 0.563 & \textbf{2.183}  \\
\textbf{\texttt{dirichlet}} & \textbf{88.6\%} & \textbf{92.0\%} & 85.2\% & \textbf{0.204} & 0.636 & \textbf{0.559} & 2.369  \\
\Xhline{0.5\arrayrulewidth}
\multicolumn{8}{l}{\textbf{MTSD/border}} \\
\textbf{} & $\mathrm{acc}$ & $\mathrm{prec}_{\cs}$ & $\mathrm{rec}_{\cs}$ & $\langle D \rangle$ & $\langle D\rangle_{\widetilde{w}}$ & $\langle H \rangle$ & $\langle H\rangle_{\widetilde{w}}$ \\
\Xhline{0.25\arrayrulewidth}
\textbf{\texttt{hard}} & 77.6\% & \textbf{79.1\%} & 90.2\% & 0.280 & 0.278 & 0.736 & 0.735  \\
\textbf{\texttt{soft}} & 78.1\% & 78.1\% & \textbf{90.4\%} & 0.263 & 0.261 & \textbf{0.664} & \textbf{0.662}  \\
\textbf{\texttt{dirichlet}} & \textbf{79.8\%} & 78.5\% & 89.2\% & \textbf{0.259} & \textbf{0.258} & 0.673 & 0.671  \\
\Xhline{0.5\arrayrulewidth}
\multicolumn{8}{l}{\textbf{MTSD/shape}} \\
\textbf{} & $\mathrm{acc}$ & $\mathrm{prec}_{\cs}$ & $\mathrm{rec}_{\cs}$ & $\langle D \rangle$ & $\langle D\rangle_{\widetilde{w}}$ & $\langle H \rangle$ & $\langle H\rangle_{\widetilde{w}}$ \\
\Xhline{0.25\arrayrulewidth}
\textbf{\texttt{hard}} & 93.3\% & 50.6\% & 67.2\% & \textbf{0.121} & 0.253 & 0.420 & 0.965  \\
\textbf{\texttt{soft}} & 93.6\% & 48.5\% & \textbf{75.0\%} & 0.128 & \textbf{0.227} & \textbf{0.317} & \textbf{0.648}  \\
\textbf{\texttt{dirichlet}} & \textbf{94.1\%} & \textbf{58.3\%} & 65.6\% & 0.129 & 0.245 & 0.337 & 0.697  \\
\Xhline{0.5\arrayrulewidth}
\multicolumn{8}{l}{\textbf{MTSD/side}} \\
\textbf{} & $\mathrm{acc}$ & $\mathrm{prec}_{\cs}$ & $\mathrm{rec}_{\cs}$ & $\langle D \rangle$ & $\langle D\rangle_{\widetilde{w}}$ & $\langle H \rangle$ & $\langle H\rangle_{\widetilde{w}}$ \\
\Xhline{0.25\arrayrulewidth}
\textbf{\texttt{hard}} & 81.3\% & 26.8\% & 54.3\% & 0.264 & 0.495 & 0.761 & 1.616  \\
\textbf{\texttt{soft}} & 82.7\% & 31.4\% & \textbf{62.9\%} & 0.228 & \textbf{0.371} & 0.576 & \textbf{1.103}  \\
\textbf{\texttt{dirichlet}} & \textbf{83.6\%} & \textbf{40.8\%} & 57.1\% & \textbf{0.211} & 0.376 & \textbf{0.562} & 1.120  \\
\end{tabularx}
\caption{Comparison of the proposed Dirichlet model to both baseline models on all test datasets and attributes. We evaluate the models on hard metrics such as accuracy, (\cs) precision and (\cs) recall, and on soft metrics, which include mean soft label distance, as defined in equation \eqref{eq:def_softdist}, and mean cross entropy. For the hard metrics, larger values are better than smaller ones; for the soft label distance and cross entropy, the reverse is true. For each dataset and each metric, the best value is highlighted in bold.}
\label{tab:metrics}
\end{table}

We first compare our Dirichlet model against the two baselines on all test datasets and attributes in terms of selected metrics. We consider both \textit{hard metrics}, which compare discrete labels, and \textit{soft metrics}, which quantify discrepancies between predicted soft labels and the true response distributions. In the case of the Dirichlet model, we reduce the predicted Dirichlet distribution to a point estimate as presented in equation \eqref{eq:posterior_mode}. In the following, let $\hat{y}_t$ denote the label predicted for a task $t$ by one of the models and $y_t$ the reference label, i.e.\ the actual majority response provided by the crowd. For the hard label comparison, we then consider the following metrics:
\begin{itemize}
    \item Accuracy $\mathrm{acc} = \abs{\mathcal{T}^{\mathrm{test}}}^{-1} \sum_{t\in\mathcal{T}^{\mathrm{test}}} \pmb{1}\left[\hat{y}_t = y_t\right]$,
    \item \englanfz{can't solve} precision $\left. \mathrm{prec}_{\cs} = \sum_t \pmb{1}\left[\hat{y}_t = \cs, y_t = \cs\right] \middle/ \sum_t \pmb{1}\left[\hat{y}_t = \cs \right] \right.$,
    \item \englanfz{can't solve} recall $\left. \mathrm{rec}_{\cs} = \sum_t \pmb{1}\left[\hat{y}_t = \cs, y_t = \cs\right] \middle/ \sum_t \pmb{1}\left[y_t = \cs \right] \right.$.
\end{itemize}
The choice of hard metrics reflects our asymmetrical interest in the tasks. On the one hand, we are interested in automating large parts of (potentially trivial) tasks, which requires optimizing accuracy. On the other hand, we investigate more nuanced features of the models, here the ability to correctly identify unsolvable tasks. For the soft label comparison we consider
\begin{itemize}
    \item average soft label distance $\langle D \rangle = \abs{\mathcal{T}^{\mathrm{test}}}^{-1} \sum_{t\in\mathcal{T}^{\mathrm{test}}} D\left( \hat{\vb{q}}_t, \vb{q}_t \right)$ and its reweighted counterpart $\langle D \rangle_{\widetilde{w}}$,
    \item average cross-entropy $\langle \mathrm{H}  \rangle = \abs{\mathcal{T}^{\mathrm{test}}}^{-1} \sum_t \mathrm{H}( \vb{q}_t \mid \hat{\vb{q}}_t)$ and its reweighted counterpart $\langle \mathrm{H} \rangle_{\widetilde{w}}$. Where ${\mathrm{H}( \vb{q} \mid \hat{\vb{q}}) = -\sum_k q^k \log \hat{q}^k}$ denotes the usual cross-entropy between the real soft label $\vb{q}$ and the model prediction $\hat{\vb{q}}$.
\end{itemize}
We include cross-entropy alongside soft label distance in our comparison because we think it is common and familiar to the reader.
For both soft label metrics, we consider a weighted variant in which each data point is included in the averaging according to the soft label weight described in equation \eqref{eq:soft_label_weights}. This allows us to look at the data in a way that addresses the unbalanced nature of the data. Generally speaking, we expect to be able to make a more refined comparison of the models using the weighted metrics. However, we emphasize that for the purpose of auto-annotation, in many cases it is not necessary to correctly solve these more difficult, underrepresented cases, as long as the model predictions are appropriately thresholded, as described in Section \ref{sec:automated_annotation}.

\paragraph{Results.}
In order to reduce statistical fluctuation of the model realizations, we repeat the training of each of the three models five times and then calculate the geometric median of the predictions. The results of our comparison of the proposed Dirichlet model to the models \texttt{baseline\_hard} and \texttt{baseline\_soft} are presented in Table \ref{tab:metrics}.
In light of these results, we would like to make a few observations. 

\textbf{(i)} The Dirichlet model generally achieves better results than both baselines in terms of (hard) accuracy and (\cs) precision, for some attributes even significantly. For (\cs) recall, the results are less stable and the Dirichlet model tends to lag behind the two baseline models. We conclude that the Dirichlet model is more conservative in predicting the unsolvability of a task. However, it is notable that a model trained to predict soft labels outperforms the hard label model on a hard metric like accuracy, especially since the hard label model was trained on already aggregated responses rather than individual annotator responses.
\textbf{(ii)} As expected, both soft label models outperform the hard label learner in almost all cases regarding soft metrics. It is difficult to rank the Dirichlet model and the soft baseline, but they both reconstruct crowd worker response distributions similarly well, with a potential small advantage for the Dirichlet model. The Dirichlet model's additional information, namely the number of responses needed to calculate the posterior distribution, potentially further regularizes the training. Thus, the mode of the predicted posterior distribution generally provides an estimate of the actual response distribution that is as good as or better than a model trained directly to predict that discrete distribution.
\textbf{(iii)} The columns for the weighted variants of the soft metrics indicate that the models sometimes make significantly worse predictions within underrepresented parts of the data compared to more regular instances. This effect is most pronounced for dichotomous questions with very few positive cases (such as the occlusion questions on MTSD). We assume that certain parts of the data in such datasets cannot be automated or can only be automated at a great expense of additional data and better training protocols. 
However, the goal of the framework is not to automate all tasks, but only those identified as particularly simple. This will be discussed in the next section.

All models have essentially the same architecture and the same number of model parameters (except for a small negligible number of additional $(C+1)^2$ parameters in the case of the Dirichlet model).
Therefore, all differences within the table, except for the stochasticity of training, are due to better use of data and smarter training objectives. It is also found that soft learners are easier to train than the hard baseline, which is more dependent on the weighting of the losses. Two justified rules of thumb are: \textbf{(1)} \textit{training on soft labels yields better models than training on hard ones}, and \textbf{(2)} \textit{models informed by absolute response distributions potentially perform better than pure soft label learners}.
\subsection{Automated Annotation}
\label{sec:automated_annotation}

\begin{figure}
    \centering
    \includegraphics[width=\textwidth]{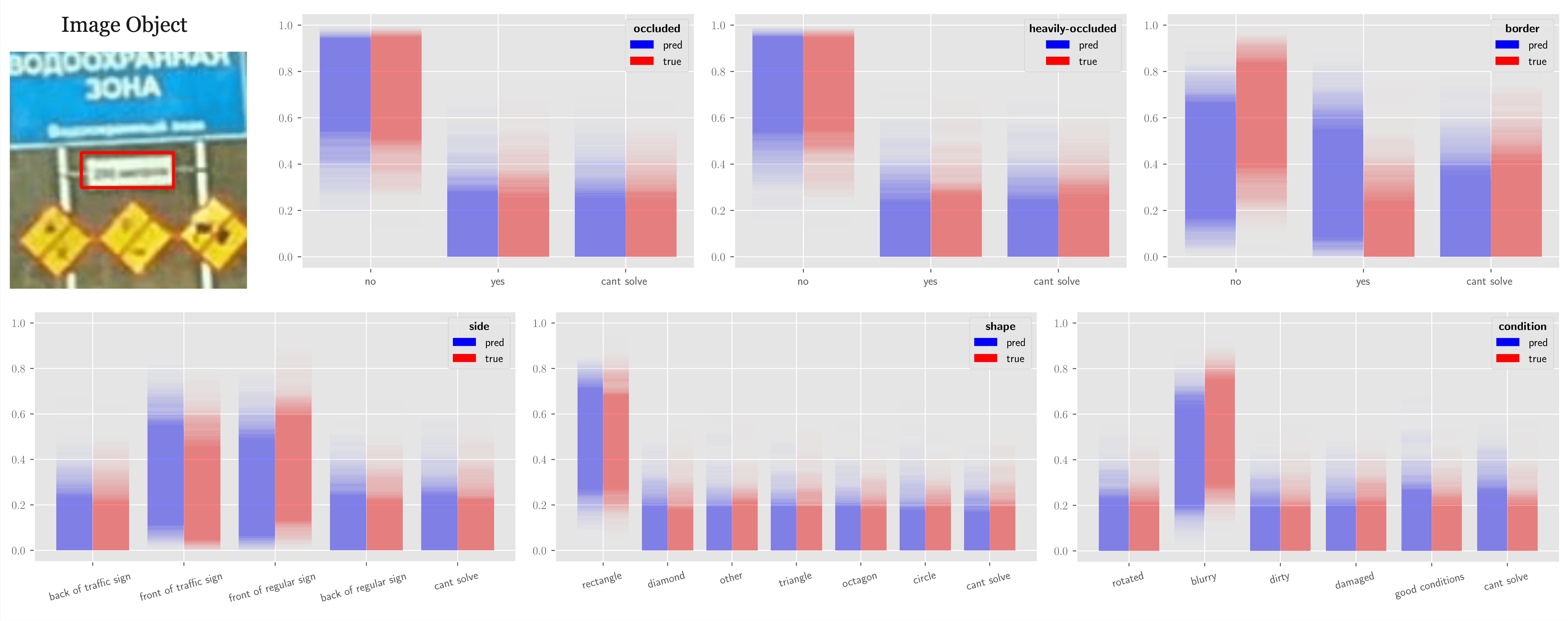}
    \caption{Example predictions of the Dirichlet model for all six attributes of one object of the Mapillary test dataset (top left). As a visualization of the predicted Dirichlet distributions, we show soft bar charts that represent repeatedly sampled realizations of possible response vectors.}
    \label{fig:butternut_example}
\end{figure}

In the case of the baseline models, predictions can be interpreted as bar charts that identify one or more categories as the most likely to be correct. Predictions of the Dirichlet model, on the other hand, correspond to Dirichlet distributions, i.e.\ distributions over probability vectors. Such distributions can be visualized, e.g.\ by probabilistic (or soft) bar charts, as shown in Figure \ref{fig:butternut_example} for all attributes of the Mapillary dataset for an example object. A single such probability vector, the mode of the respective Dirichlet distribution, can be chosen as the point estimator for the purpose of auto-annotation. We compare the predictions obtained in this way from the proposed Dirichlet model with those from the hard label model.

\begin{figure}
    \centering
    \floatbox[{\capbeside\thisfloatsetup{capbesideposition={right,top},capbesidewidth=0.4\textwidth}}]{figure}[\FBwidth]
    {\caption{Automation-correctness curve for the \texttt{ECP/human-being} test dataset. Shown is a comparison of the hard label model (blue) and the proposed Dirichlet model (red). The shaded areas indicate approximate $95\%$ confidence intervals calculated from the sample quantiles over $B=1024$ bootstrap samples. The curve is used for out of sample predictions of how correct the auto-annotation model is when a certain level of automation is selected. Each point on the curve corresponds to a threshold parameter on the model confidence. The choice of threshold determines how many of the automatically annotated instances we trust and which part of the data should be evaluated by human annotators.}\label{fig:ecp_automation_correctness}}
    {\includegraphics[width=0.55\textwidth]{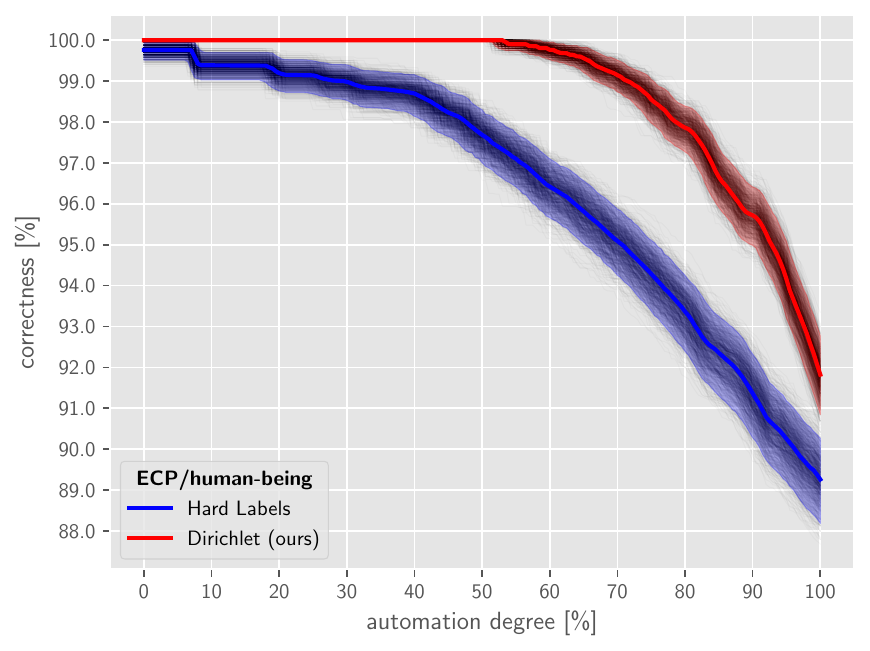}}
\end{figure}

\begin{figure}
    \centering
    \includegraphics[width=\textwidth]{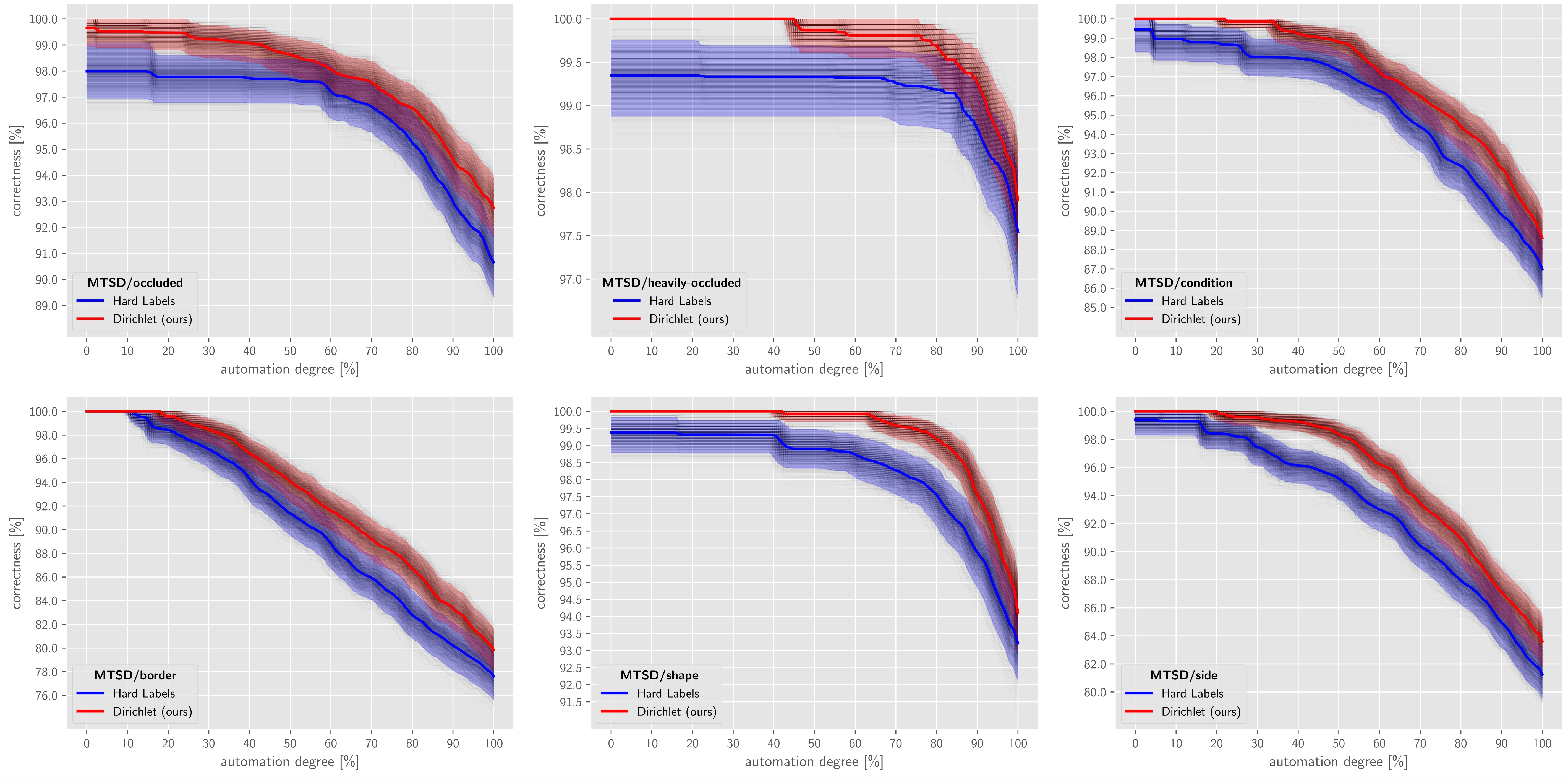}
    \caption{Automation-correctness curves for each of the six attributes of the \texttt{MTSD} objects. For each attribute, we show the accuracy achieved at the respective degree of automation for the baseline model (blue) and the Dirichlet model (red). For each dataset and model, we calculate approximate $95\%$ confidence intervals (shaded areas) obtained from $B=1024$ bootstrap samples.}
    \label{fig:butternut_automation_correctness}
\end{figure}

\begin{table}[ht]
\centering
\footnotesize
\begin{tabularx}{\textwidth}{XXX}
\Xhline{0.25\arrayrulewidth}
 & Automation $\mathrm{CI}_{95\%}$ & Accuracy $\mathrm{CI}_{95\%}$ \\
\Xhline{0.25\arrayrulewidth}
ECP/human-being & (68.8\%, 78.8\%) & (97.8\%, 99.4\%) \\
MTSD/occluded & (55.5\%, 68.9\%) & (97.3\%, 99.0\%) \\
MTSD/heavily-occluded & (86.8\%, 94.1\%) & (98.8\%, 99.7\%) \\
MTSD/condition & (2.0\%, 51.6\%) & (98.5\%, 100.0\%) \\
MTSD/border & (26.1\%, 34.7\%) & (97.6\%, 99.6\%) \\
MTSD/shape & (75.6\%, 86.7\%) & (98.6\%, 99.7\%) \\
MTSD/side & (35.0\%, 52.7\%) & (98.3\%, 99.8\%) \\
\end{tabularx}
\caption{Evaluation of the $99\%$ target accuracy threshold from validation to test datasets. First, we calculate an automation-correctness curve, as shown previously, on a validation dataset by selecting a point on the curve that provides the highest degree of automation with a model correctness of no less than $99\%$. To obtain an unbiased prediction of the out-of-sample performance of the model, we evaluate the selected threshold on the automation-correctness curve corresponding to a retained test dataset. We expect that for the selected threshold, the model accuracy on the test dataset will also be around $99\%$. By bootstrapping on the validation dataset, we obtain not one, but a whole set of thresholds. From this distribution of thresholds, we obtain approximate $95\%$ confidence intervals for both the degree of automation and model accuracy on the test dataset.}
\label{tab:automation_accuracy_intervals}
\end{table}

Our understanding of auto-annotation is detailed below. For each annotated dataset, such as the validation or test subset, we repeatedly calculate realizations of a curve through resampling. These curves emerge by selecting successive thresholds $0 < c < 1$ and reducing the analyzed dataset to the tasks $t$ for which the model's predicted confidence (according to equation \eqref{eq:def_confidence}) exceeds the threshold for the selected annotation attribute. For each threshold, we determine the proportion of the reduced dataset within the total dataset, and the accuracy achieved for this subset. Each threshold $c$ thus corresponds to a point representing the accuracy for a given proportion of the target dataset. The collection of these points forms a parameterized curve, referred to as the \textit{automation-correctness curve}. By bootstrapping the dataset, we can calculate a set of curves, from which we derive approximate confidence intervals. This curve (or set of curves) allows us to estimate the degree of automation expected to meet a specified quality target (expressed as accuracy).
Initially, choosing accuracy as the target metric might seem counterintuitive due to the data's unbalanced nature. However, the goal is to identify the point that provides sufficiently high accuracy (i.e., $(1-\varepsilon)100\%$ for a small $\varepsilon \ll 1$), allowing us to auto-annotate the trivial parts of the data that do not require human intervention.

Figure \ref{fig:ecp_automation_correctness} shows the corresponding curves for the ECP dataset, and Figure \ref{fig:butternut_automation_correctness} shows them for the Mapillary dataset, each based on $B=1024$ bootstrap samples. The results for the hard label model (blue) and the Dirichlet model (red) are presented. The bold lines indicate the median curve, while the shaded areas represent $95\%$ confidence intervals based on sample quantiles. In all cases, the Dirichlet model significantly outperforms the hard label model, with smaller bootstrap uncertainties. For instance, with the ECP dataset, more than half of the data could be automatically annotated, reflecting the asymmetric distribution of relevant objects. This method allows us to identify trivial cases, directing human labelers' efforts to more difficult examples.

A pertinent question is how to determine thresholds for automatic annotation based on these curves. Using the test data curve to set thresholds risks double dipping\footnote{By \textit{double dipping} we mean the practice of using the test dataset both for the selection of a threshold and for the evaluation of the model under the selected threshold. We propose to use the test dataset exclusively for the final out-of-sample evaluation of the model and to choose the threshold in a programmatic way independently of this.}. Instead, we propose determining the threshold using the validation dataset, which is spent little or not at all on tuning other hyperparameters. We repeatedly compute realizations of the automation-correctness curve on the validation dataset. For each curve and a target accuracy (here, $99\%$), we find the smallest threshold that achieves at least the target accuracy. These thresholds are then evaluated on the test dataset curves, allowing us to derive joint confidence regions for automation and correctness. For simplicity, we marginalize these joint confidence regions. Table \ref{tab:automation_accuracy_intervals} shows the results at a $95\%$ confidence level for all datasets. Ideally, the confidence interval for correctness includes the target accuracy of $99\%$, which is achieved for each dataset. These estimates indicate that large portions of the data can be auto-annotated, with moderate intervals except for the \texttt{condition} attribute in the Mapillary dataset. Users can adjust the threshold based on their willingness to trade model quality for higher automation. For simplicity, we use the median of the realized thresholds in further experiments and examine the remaining parts of the datasets after thresholding at these chosen values.

\subsection{Predicting Ambiguity}

\begin{figure}
    \centering
    \includegraphics[width=\textwidth]{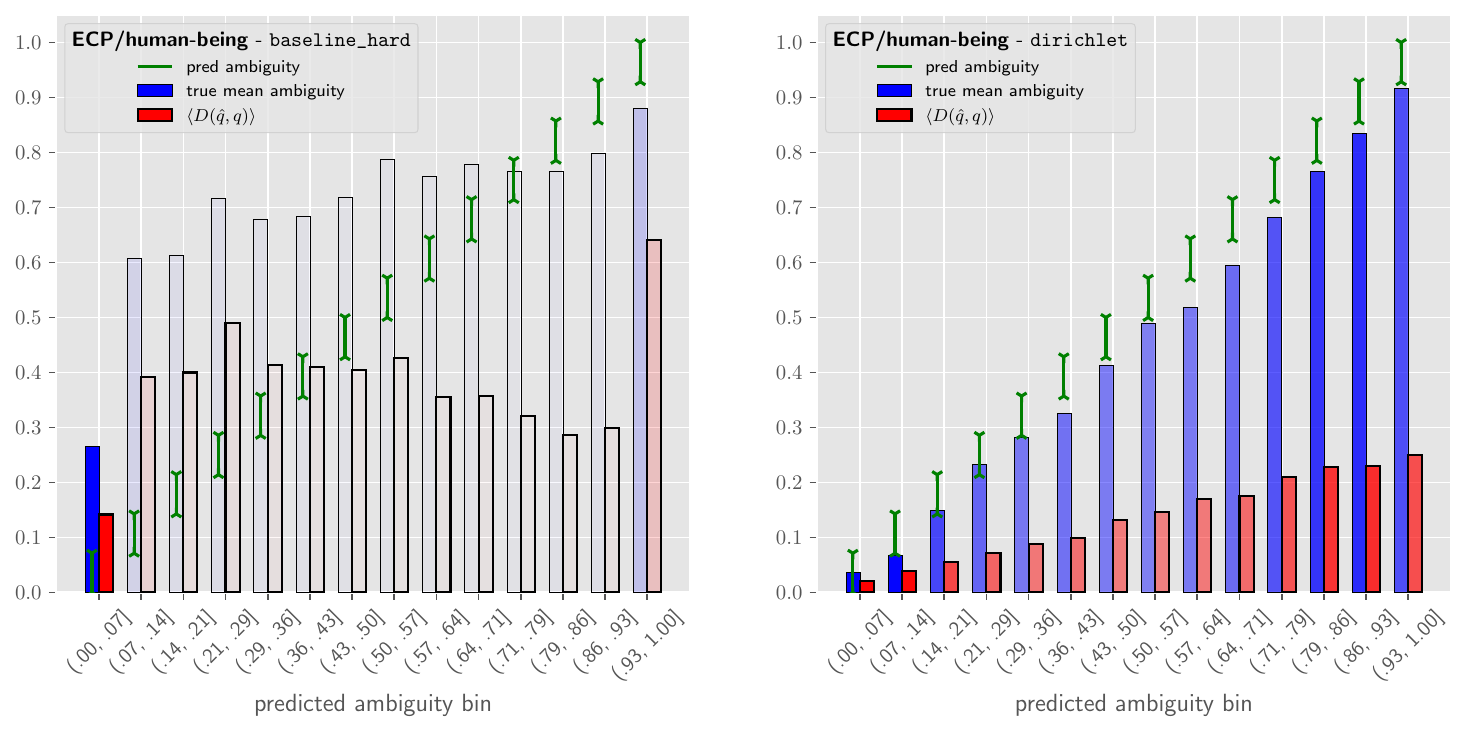}
    \caption{Calibration of the ambiguity predicted by the hard label model (left) and the Dirichlet model (right) on the \texttt{ECP/human-being} test dataset. Objects are first distributed over equidistant bins according to the predicted ambiguity. Within each bin, the actual mean ambiguity is determined (blue), as well as the mean probabilistic distance (red).
    The transparency of the bars is inversely proportional to the number of objects in the respective bin.
    In the case of the Dirichlet model, it can be seen that the predictions generally overestimate the true ambiguity, but do so in a very linear and predictable way. The model trained on hard labels captures label ambiguity only weakly and much less systematically.}
    \label{fig:amb-histogram}
\end{figure}

\begin{figure}[ht]
    \centering
    \includegraphics[width=0.975\textwidth]{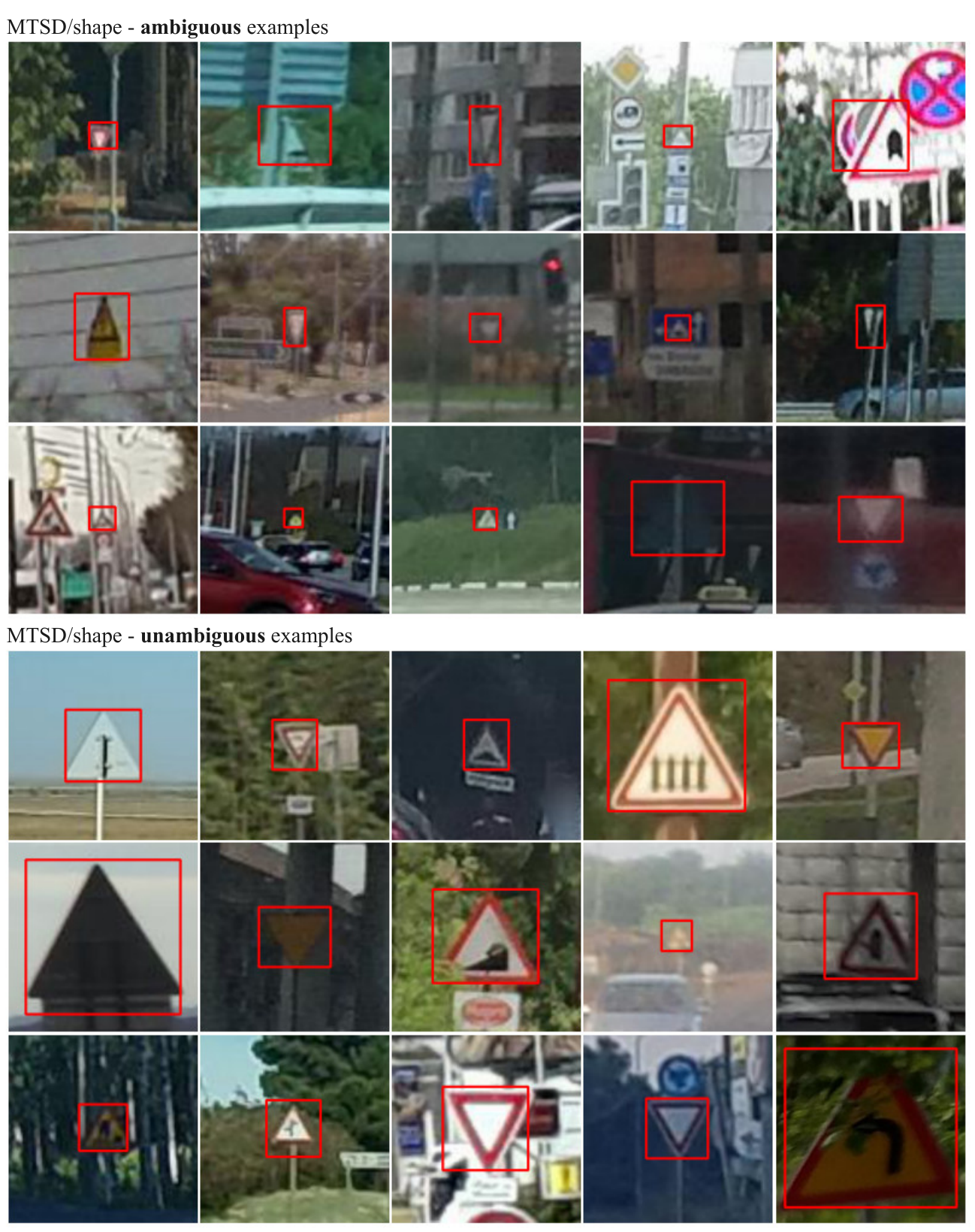}
    \caption{Examples of traffic signs with high ambiguity (top) vs.\ low ambiguity (bottom) regarding the question of the shape of the objects. All signs are rated as triangular according to the majority vote, but the ambiguity of the answers given by humans is greater for the upper examples, e.g.\ due to a large distance of the objects to the camera, the orientation of the signs, occlusion or inadequately drawn bounding boxes.}
    \label{fig:ambiguous-vs-unambiguous}
\end{figure}

A different use case for auto-annotation, beyond what has been discussed so far, is to automatically assess the difficulty or ambiguity of an annotation task. Ambiguity, as introduced in Equation \eqref{eq:def-ambiguity}, refers to a property of the probability vector associated with a task.
Taking into account the ambiguity of the labels assigned by humans allows a machine to make predictions without having to make a hard, possibly inadequate choice regarding a single category.
Human uncertainty in annotation tasks can have various causes such as intrinsic difficulties associated with the task design, task input or limited suitability of the annotators solving the tasks. The proposed ambiguity measure can provide an indication that certain outputs of an annotation task are associated with such possible difficulties, without disambiguating the causes in detail.
It is important to mention that ambiguity, as we use it here, encompasses both \textit{aleatoric} and \textit{epistemic} uncertainty. Epistemic uncertainty arises when a task is often underspecified, meaning it cannot be solved or can only be solved incompletely due to the way the annotation question or the object to be annotated is presented. As a result, annotators may be unable to provide a clear answer or may resort to the \englanfz{can't solve} option. Conversely, the ambiguity measure also captures aleatory uncertainty, which is intrinsic and irreducible. This type of uncertainty stems from the inherent unpredictability of human behavior, making the outcome of an annotation task necessarily random even under optimal task design.
Examples of such objects that are not assigned a clear label by human annotators with respect to the \texttt{MTSD/shape} question can be seen in the upper half of Figure \ref{fig:ambiguous-vs-unambiguous}. On the other hand, the lower half of the figure shows examples for which the annotators can clearly identify triangular traffic signs.

Using the \texttt{ECP/human-being} task, we find that soft label learning, particularly with the Dirichlet model, is capable of making useful predictions about the actual ambiguity of an object in relation to a task. In Figure \ref{fig:amb-histogram}, we compare predicted versus actual ambiguity for both the hard label model and the Dirichlet model. The predicted ambiguities over the test dataset are first binned equidistantly, and then the mean ambiguity of the instances within each bin is calculated.
The figure illustrates that the Dirichlet model, unlike the hard label model, can predict object ambiguity plausibly and consistently, even though the Dirichlet model tends to slightly overestimate actual ambiguity in general. For each bin, we also present the mean probabilistic distance between the prediction and the actual distributions. The graph shows that predicting distributions associated with high ambiguity is generally more challenging.

\begin{figure}[ht]
    \centering
    \includegraphics[width=0.99\textwidth]{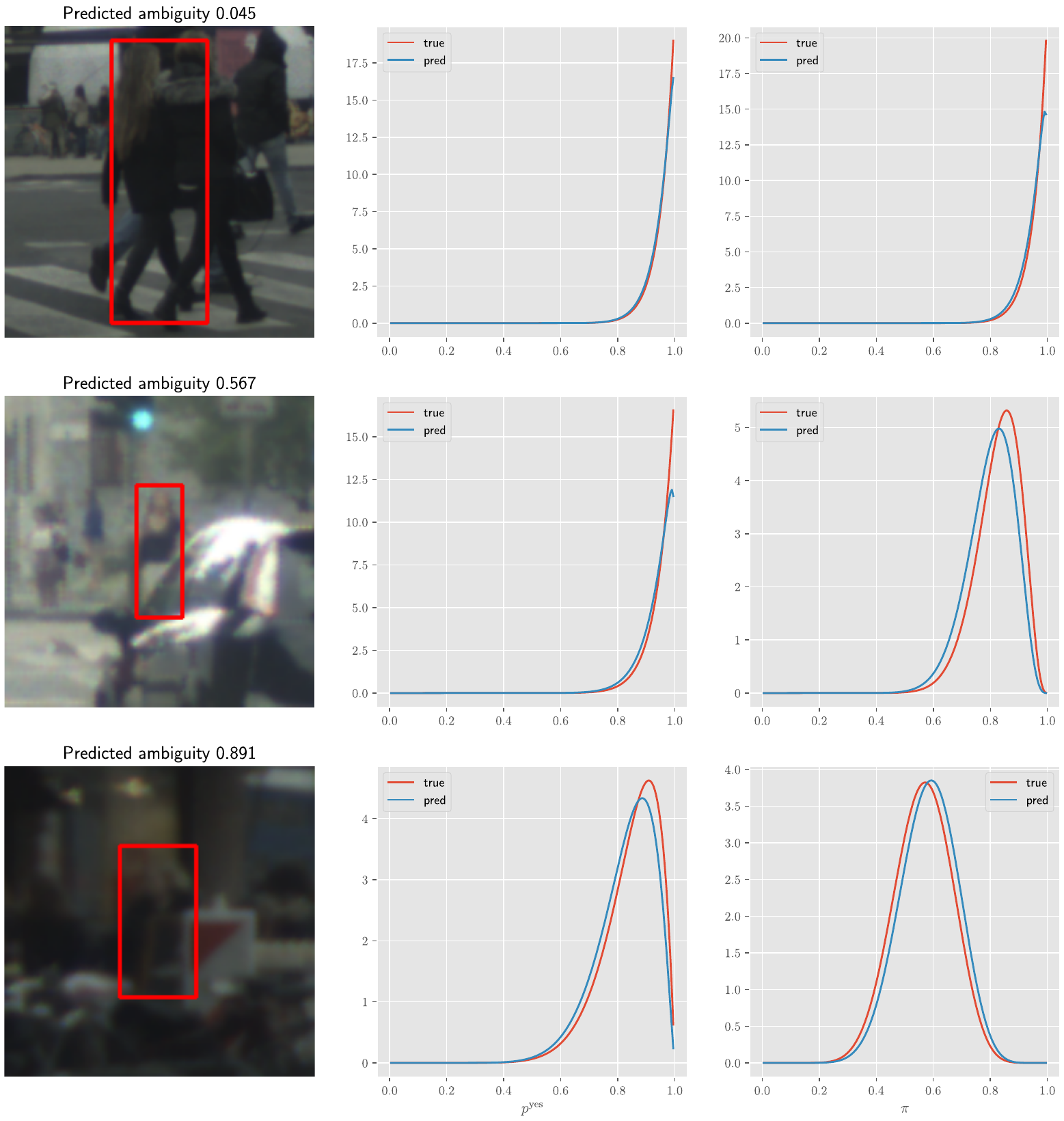}
    \caption{Three example predictions of the Dirichlet model for the \texttt{ECP/human-being} dataset: example of low ambiguity (top), medium ambiguity (middle) and high ambiguity (bottom). For each object, the predicted and the actual distribution are shown, represented by the marginal distributions for probability of success (middle column) and solvability (right column).}
    \label{fig:ecp-amb-examples}
\end{figure}

\begin{figure}[ht]
    \centering
    \includegraphics[width=0.99\textwidth]{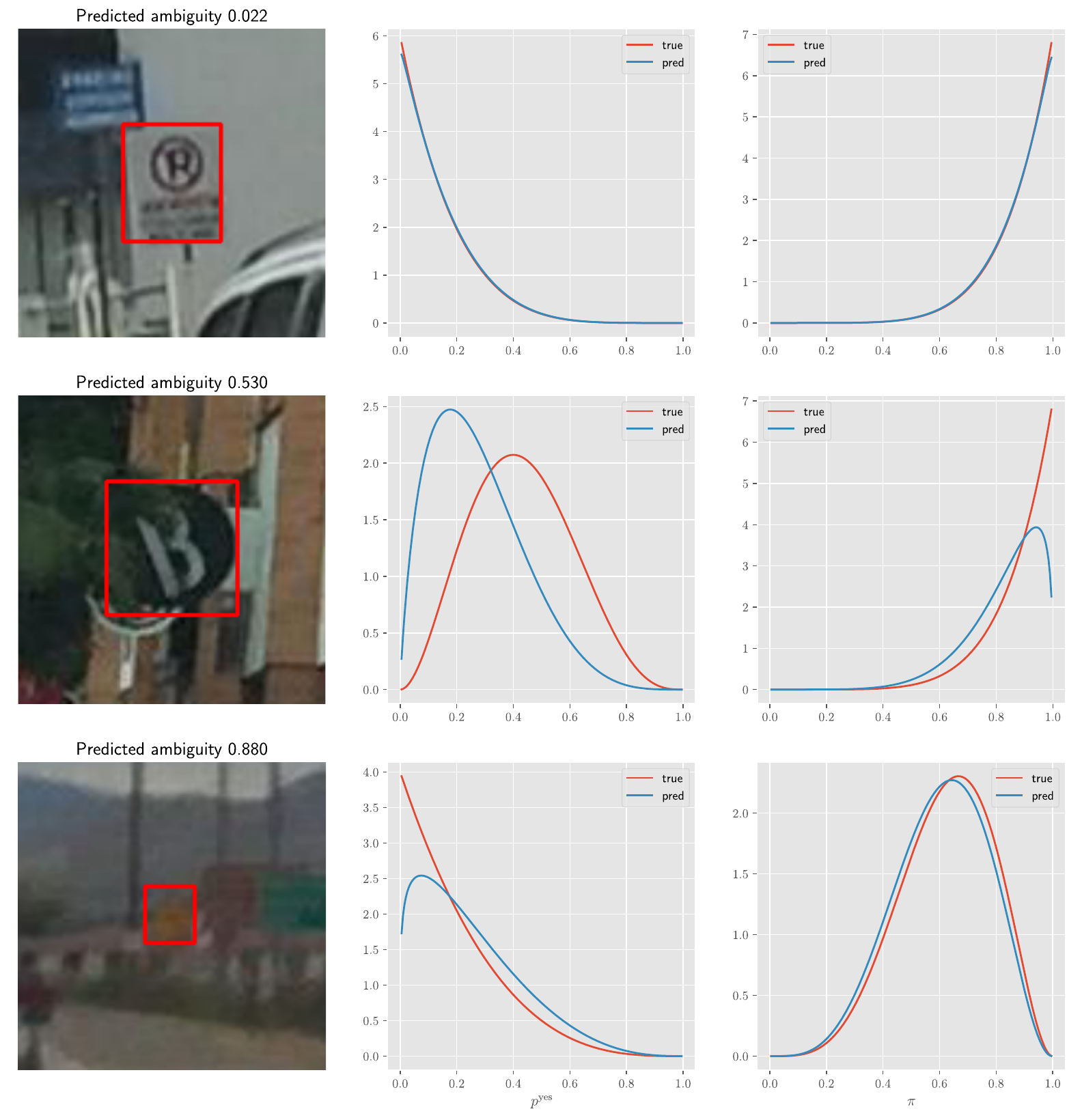}
    \caption{Three example predictions of the Dirichlet model for the \texttt{MTSD/heavily-occluded} dataset: example of low ambiguity (top), medium ambiguity (middle) and high ambiguity (bottom). For each object, the predicted and the actual distribution are shown, represented by the marginal distributions for probability of success (middle column) and solvability (right column).}
    \label{fig:heavily-amb-examples}
\end{figure}

The model's capacity to predict ambiguity provides a way to filter datasets for interesting content in a highly effective way. Examples of tasks predicted as low, medium and high ambiguity are shown for the \texttt{ECP/human-being} dataset in Figure \ref{fig:ecp-amb-examples}. For each image object, we show the model predictions and actual posterior distributions in the form of the marginalized distributions of success probability and solvability, as introduced in equations \eqref{eq:transformed_dirichlet_pi} and \eqref{eq:transformed_dirichlet}, respectively. These few examples give a qualitative impression of the typical cases associated with each ambiguity level: clearly recognizable pedestrians in the case of low ambiguity, strongly occluded pedestrians in the case of medium ambiguity, and barely recognizable objects in the case of high ambiguity. A similar qualitative behavior can also be identified in the context of the other datasets, as shown in the example of the \texttt{MTSD/heavily-occluded} data in Figure \ref{fig:heavily-amb-examples}.

\subsection{Prediction as Prior}

We now explore how the proposed Dirichlet model supports annotating examples that cannot be automatically annotated with certainty. This addresses the third point in our introduction: our model's integration into a human-driven truth inference process within a Bayesian framework. This integration is facilitated by our approach of learning distributions of soft labels instead of a single point estimate.
Specifically, we suggest replacing the uniform prior used in the data-generating process in Section \ref{sec:dirichlet-multinomial-model} with the Dirichlet model's predictions for each task. We investigate whether this approach allows for more reliable predictions of a task's underlying probability vector based on human responses.

In our \textit{repeats analysis}, we draw answers for each object (without replacement) from the observed responses per task, iteratively updating the Dirichlet distributions via Bayesian updating. Each turn uses the previous turn's posterior distribution as the new prior, which is updated with each additional response. After each update, we determine the mode of the resulting Dirichlet distribution, reducing it to a point estimate, or soft label. We compare the soft labels obtained by initializing the process with either uniform or learned priors, computing probabilistic distances after each step. By averaging these distances for each object, we obtain distributions of distances across all objects in the dataset after each step. Note that we only consider objects that remain after the thresholding from Section \ref{sec:automated_annotation}, excluding those that are trivially automatable.
A key point is how the Dirichlet model's predictions are used as priors. The predicted Dirichlet distributions are influenced by the number of responses used to infer the approximated posterior distribution. Using these distributions directly would bias the inference before observing any responses. To mitigate this, we rescale the model's predictions for any number of responses, as shown in Equation \eqref{eq:dirichlet_scaling}. We start with a base prediction, choosing $n=0$ as the number of target responses. To further regularize the influence of the prediction and reduce bias, we blend the prediction with the uniform prior, a parameter vector of ones. This convex combination determines the influence of the learned prior relative to the uninformed prior. We treat the model prediction as a \englanfz{perturbation} of the uninformative prior, blending the two distributions in a $2\mathbin{:} 1$ ratio. Thus, the machine-informed prior for each task becomes $(2/3)\pmb{1}+(1/3)\hat{\pmb{\alpha}}\vert_{n=0}$.

\paragraph{Results.}

\begin{figure}[ht]
    \centering
    \floatbox[{\capbeside\thisfloatsetup{capbesideposition={right,top},capbesidewidth=0.4\textwidth}}]{figure}[\FBwidth]
    {\caption{Inference with uniform and model-informed priors on the \texttt{ECP/human-being} dataset. We repeatedly draw answers (without replacement) from the set of actual responses observed per task and iteratively update the Dirichlet distributions. We reduce each Dirichlet distribution to a single soft label and calculate the probabilistic distance to the actual soft label given by the workers. After each step, we plot the distributions of the realized probabilistic distances. Thick lines represent the interquartile range (IQR), bold dots indicate medians, and thin extensions show the $95\%$ data range. The inference with the informed prior generally leads to a faster approximation of the observed soft labels, resulting in significantly faster convergence towards the real distribution.}\label{fig:ecp_repeat_analysis}}
    {\includegraphics[width=0.58\textwidth]{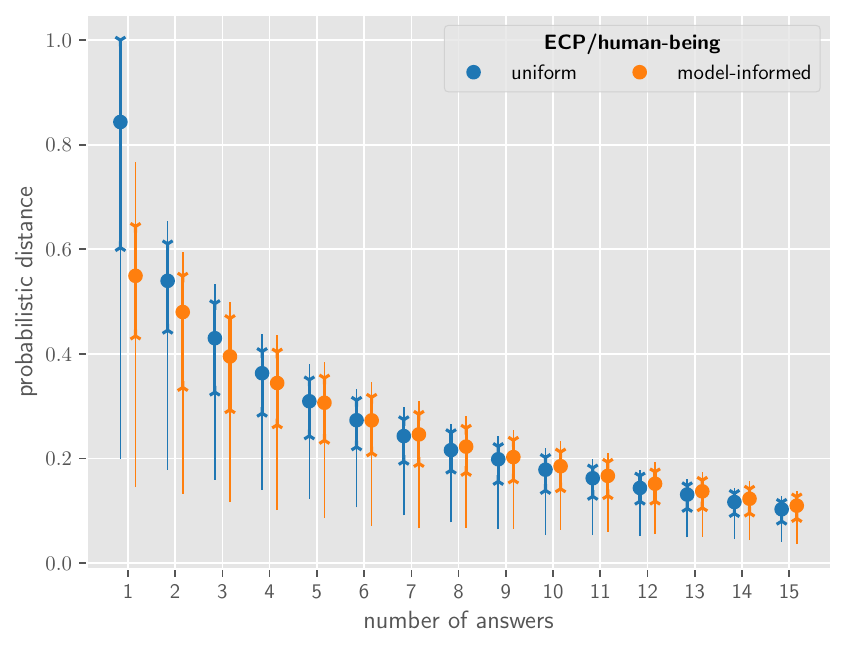}}
\end{figure}

\begin{figure}[ht]
    \centering
    \includegraphics[width=\textwidth]{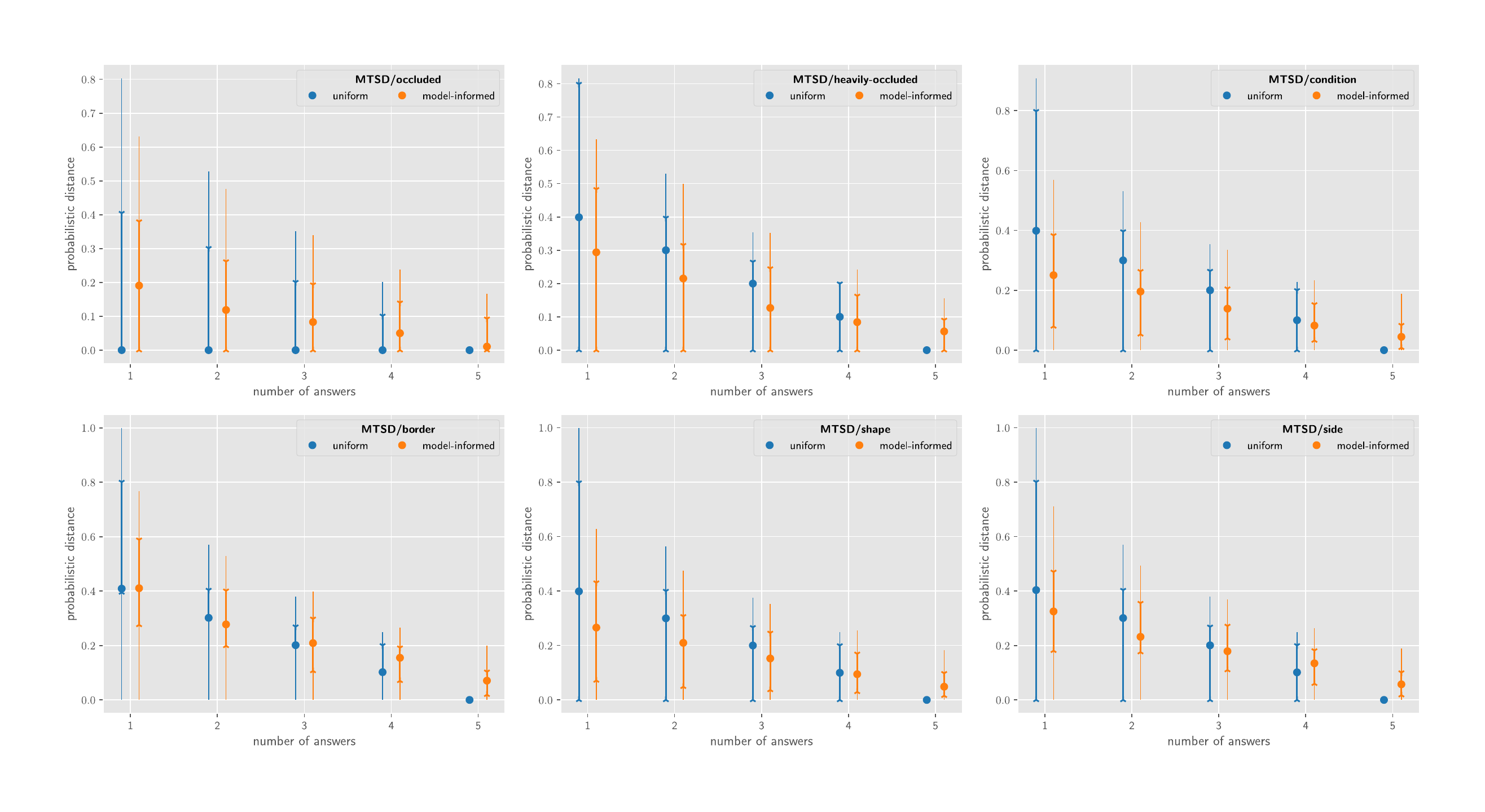}
    \caption{Inference with uniform and model-informed priors for each of the six attributes of the Mapillary data. The plot is to be understood in the same way as figure \ref{fig:ecp_repeat_analysis}. That is, each line provides information about the distribution of probabilistic distances by indicating the IQR, a $95\%$ data range and the position of the median. The model-informed prior biases the inference in a data-driven way and generally allows to approximate the actual soft labels faster with lower variance. The fact that the probabilistic distance in the case of a uniform prior is guaranteed to be zero after five observed labels is a mathematical consequence of our evaluation.}
    \label{fig:butternut_repeat_analysis}
\end{figure}

We present the results of our proposed repeats analysis in Figure \ref{fig:ecp_repeat_analysis} for the ECP dataset and Figure \ref{fig:butternut_repeat_analysis} for the Mapillary dataset. For the ECP dataset, we draw $15$ answers, while for the Mapillary dataset, we choose the maximum possible number of five responses. Each plot shows the number of responses drawn on the $x$-axis, with the $y$-axis representing the distributions of probabilistic distances across the objects in the respective dataset. The distributions are depicted similarly to boxplots: thick lines indicate interquartile ranges (IQR), with extensions to the $2.5\%$ and $97.5\%$ quantiles, and dots marking the medians.

The inference with the informed prior generally leads to a faster approximation of the observed soft labels, resulting in significantly lower probabilistic distances. For example, in the \texttt{ECP/human-being} dataset, after observing one response, around $50\%$ of the objects have a probabilistic distance greater than $0.6$. In contrast, with the learned prior, almost $75\%$ of the distances are less than $0.6$. The informed prior shows similarly strong performance across the other datasets. The exception is the \texttt{MTSD/occluded} task, where about half the data can be nearly perfectly annotated even with a small number of responses, likely due to the low prevalence of actually occluded road signs.

As we increase the number of responses, the influence of the prior diminishes, indicating that the risk of systematic errors based on the learned prior is minimal. The model biases the predictions in a positive way in the presence of fewer responses, while the influence on the inference result becomes smaller the more responses are observed.
In the case of the naive prior, reaching a probabilistic distance of zero for the maximum number of repeats is a mathematical consequence of our sampling without replacement.

\section{Conclusion}

In this paper, we present a predictive model that is capable of partly automating annotation processes and guiding truth inference.
Our model differs from hard label models in that it is not trained to predict a single hard label, but to represent the sentiment of human labelers.
Moreover, it deviates from the direct learning of soft labels by explicitly addressing the inherent uncertainty associated with soft labels, viewed as outcomes of a data-generating process in the sense of Bayesian inference.
We illustrate how the product of Bayesian inference, the posterior distribution over soft labels, is used to train a model that achieves a performance similar or even superior to models trained directly on either hard or soft labels.
Using various examples of real-world datasets, we see that the proposed Dirichlet model is able to make the inspection of large amounts of visual data more cost-effective without sacrificing quality of results.
The model solves the question of whether a person can be recognized in an image section of a road traffic scene in about $70$ to $80\%$ of the cases with an accuracy of about $99\%$.
As a consequence, annotation budget can be saved considerably or spent on other tasks that require more expensive human intelligence.
Even if the model cannot fully automate a task, it can be used to generate reliable soft labels faster. Using the learned prior to guide the inference of the soft label associated with each task allows for a significant reduction in the number of responses required from human labelers.
We also see how the proposed model, while generally overestimating the true ambiguity of human responses, does so linearly and consistently enough to make the predicted ambiguity useful for identifying and sorting difficult-to-label examples. 
We see the great value of our work in effectively addressing the challenge of annotating large datasets while explicitly considering data ambiguity. We have evidence of the practical relevance of this approach, with many ideas from this work now integrated into our data inspection and annotation platform\footnote{\url{https://www.quality-match.com/}}.

\clearpage
\printbibliography[heading=subbibliography]


\end{document}